%% file: MM_MNRAS_mt_final.tex
\documentclass[useAMS,usenatbib]{mnras}
\usepackage{amsmath,graphicx,bm,amsbsy,amssymb,aas_macros,color,enumerate,lgrind,cmap,rotating,url,array,hyperref,appendix,ctable,capt-of,setspace, float, booktabs}
\usepackage{pdflscape}
\usepackage{natbib}
\input{macros.tex}

\title[Brute-Force Mapmaking with Compact Interferometers]{Brute-Force Mapmaking with Compact Interferometers: \\ A MITEoR Northern Sky Map from 128\,MHz to 175\,MHz}
\author[H.~Zheng, et al.]{H.~Zheng$^{1}$\thanks{E-mail:
		jeff.h.zheng@gmail.com}, M.~Tegmark$^{1}$, J.~S.~Dillon$^{1,2}$, A.~Liu$^{2}$\thanks{Hubble Fellow}, A.~R.~Neben$^{1}$, S.~M.~Tribiano$^{1,3,4}$
	\newauthor  R.~F.~Bradley$^{5, 6}$, V.~Buza$^{1,7}$, A.~Ewall-Wice$^{1}$, H.~Gharibyan$^{1,8}$, J.~Hickish$^{9}$, E.~Kunz$^{1}$, 
	\newauthor J.~Losh$^{1}$, A.~Lutomirski$^{1}$, E.~Morgan$^{1}$, S.~Narayanan$^{1}$, A.~Perko$^{1,8}$, D.~Rosner$^{1}$, 
	\newauthor N.~Sanchez$^{1}$, K.~Schutz$^{1,10}$, M.~Valdez$^{1,11}$, J.~Villasenor$^{1}$, H.~Yang$^{1,8}$, K.~Zarb Adami$^{3}$, 
	\newauthor I.~Zelko$^{1,7}$, K.~Zheng$^{1}$\\
	$^{1}$Dept. of Physics and MIT Kavli Institute, Massachusetts Institute of Technology, 77 Massachusetts Ave., Cambridge, MA 02139, USA\\
	$^{2}$Dept.~of Astronomy and Radio Astronomy Lab, University of California, Berkeley, CA 94720, USA\\	
	$^{3}$Science Dept.~Borough of Manhattan Community College, City University of New York, New York, NY 10007, USA\\
	$^{4}$Dept.~of Astrophysics, American Museum of Natural History, New York, NY 10024\\
	$^{5}$Dept.~of Electrical and Computer Engineering, University of Virginia, Charlottesville, VA 22904, USA\\
	$^{6}$National Radio Astronomy Observatory, Charlottesville, VA 22903, USA\\
	$^{7}$Dept.~of Physics, Harvard University, Cambridge, MA 02138, USA\\
	$^{8}$Dept.~of Physics, Stanford University, Stanford, CA 94305, USA\\
	$^{9}$Dept.~of Physics, University of Oxford, Oxford, OX1 3RH, United Kingdom\\
	$^{10}$Berkeley Center for Theoretical Physics, University of California, Berkeley, CA 94720, USA\\	
	$^{11}$Dept.~of Astronomy, Boston University, Boston, MA 02215, USA}

\date{\today}

\begin{document}

\maketitle

\begin{abstract}
	We present a new method for interferometric imaging that is ideal for the large fields of view and compact arrays common in 21~cm cosmology. We first demonstrate the method with simulations for two very different low frequency interferometers, the Murchison Widefield Array (MWA) and the MIT Epoch of Reionization (MITEoR) Experiment. We then apply the method to the MITEoR data set collected in July 2013 to obtain the first northern sky map from 128\,MHz to 175\,MHz at $\sim\!\!2^\circ$ resolution, and find an overall spectral index of $-2.73\pm0.11$. The success of this imaging method bodes well for upcoming compact redundant low-frequency arrays such as HERA. Both the MITEoR interferometric data and the 150\,MHz sky map are available at \url{http://space.mit.edu/home/tegmark/omniscope.html}.
\end{abstract} 

\begin{keywords}
	Cosmology: Early Universe -- Radio Lines: General -- Techniques: Interferometric -- Methods: Data Analysis 
\end{keywords}

%%%%%%%%%%%%%%%%%%%%%%%%%%%%%
%%%%%%%%%%%%%%%%%%%%%%%%%%%%%
%%%%%%%%%%%%%%%%%%%%%%%%%%%%%
\section{Introduction}

Mapping neutral hydrogen throughout our universe via its redshifted 21 cm line offers a unique opportunity to probe the cosmic ``dark ages'', the formation of the first luminous objects, and the Epoch of Reionization (EoR). In recent years a number of low-frequency radio interferometers designed to probe the EoR have been successfully deployed, such as the Low Frequency Array (LOFAR; \citealt{LOFAR}), the Murchison Widefield Array (MWA; \citealt{MWA}), the Donald C. Backer Precision Array for Probing the Epoch of Reionization (PAPER; \citealt{PAPER}), the 21~cm Array (21CMA; \citealt{21cma}), and the Giant Metrewave Radio Telescope (GMRT; \citealt{GMRT}). Unfortunately, the cosmological 21~cm signal is so faint that none of the current experiments have detected it yet, although increasingly stringent upper limits have recently been placed \citep{GMRT2, MWAJosh, PAPERpspec, jacobs_et_al2014, PAPERpspec2}. Fortunately, the next generation Hydrogen Epoch of Reionization Array (HERA; \citealt{HERA}) is already being commissioned and larger future arrays, such as the Square Kilometre Array (SKA; \citealt{SKA}), are in the planning stages. 

Mapping diffuse structure is important for EoR science. A major challenge in the field is that foreground contamination is perhaps four orders of magnitude larger in brightness temperature than the cosmological hydrogen signal \citep{GSM, MWAJosh, PAPERpspec, PAPERpspec2}. Many first generation experiments therefore focus on a foreground-free region of Fourier space, giving up on any sensitivity to the foreground-contaminated regions \citep{foreground3, pober_etal2013b, liu_et_al2014a, liu_et_al2014b, dillon_et_al2015}. To access those regions and regain the associated sensitivity, one must accurately model and remove such contamination, a challenge that requires even greater sensitivity as well as more accurate calibration and beam modeling than the current state-of-the-art in radio astronomy (see \citealt{FurlanettoReview,miguelreview} for reviews). Moreover, an accurate sky model is important for calibrating these low-frequency arrays. For non-redundant arrays---arrays with few or no identical baselines---such as the MWA, modeling the diffuse structure is necessary for calibrating short baselines \citep{foreground10, foreground11}. For redundant arrays such as PAPER, MITEoR, and HERA \citep{HERA}, although they can apply redundant calibration which solves for per antenna calibration gains without any sky models \citep{MITEoR, PAPERpspec2}, they do need a model of the diffuse structure to lock the overall amplitude of their measurements \citep{JacobsFluxScale}.

The Global Sky Model (GSM; \citealt{GSM}) is currently the best model available for diffuse Galactic emission at EoR frequencies. It has been widely used in the EoR community as a foreground model. The sky maps used in the GSM that are close to the EoR frequency (100\,MHz to 200\,MHz) are the 1999 MRAO+JMUAR map \citep{Alvarez1997MRAO, Maeda1999JMUAR} at 45\,MHz and $3.6^\circ$ resolution, and the 1982 Haslam map \citep{Haslam1981, Haslam1982, Remazeilles2015Haslam} at 408\,MHz and $0.8^\circ$ resolution. Other notable sky maps include the 1970 Parkes maps \citep{Landecker1970Parkes} near the equatorial plane at 85\,MHz and 150\,MHz, with $3.5^\circ$ and $2.2^\circ$ resolution, respectively. However, both the frequency and sky coverage of these maps are insufficient to achieve the precision necessary for 21\,cm cosmology.

All the maps listed above were made using steerable single dish antennas. The sensitivity necessary for EoR science requires a large collecting area, for which steerable single dish radio telescopes become prohibitively expensive \citep{tegmark_zaldarriaga2009}. As a result, the aforementioned EoR experiments have all opted for interferometry, combining a large number of independent antenna elements which are (except for GMRT) individually more affordable. Since these low-frequency instruments are designed for EoR science, they differ in many ways from traditional radio interferometers. For example, PAPER's array layout maximizes redundancy for optimal sensitivity and calibration \citep{parsons_et_al2012a}, which makes it unsuitable for standard interferometric imaging. MWA, on the other hand, has a minimally redundant layout suitable for imaging, and customized algorithms have been developed to optimize its imaging performance such as \citet{sullivan_et_al2012} and \citet{WSClean}. Some progress has been made to create maps of large-scale structure \citep{wayth_et_al2015}, they have relied on traditional algorithms designed for narrow-field imaging.

Fundamentally, interferometric imaging is a linear inversion problem, since the visibility data are linearly related to the sky temperature: the vector of measured data equals the vector of sky temperature times some huge matrix, with added noise. A number of map-making frameworks using this matrix approach have been proposed in recent years. For example, \citet{CHIMEMM} proposed a technique on spherical harmonics designed for drift scan instruments. \citet{LOFARMM} focused mostly on imaging fields of view on the scale of a few degrees. While \citet{JoshMM} develops much of the same formalism that we use, they focus less on high-fidelity mapmaking and more on propagating mapmaking effects into 21~cm power spectrum estimation. The ``brute force'' matrix inversion and full-sky imaging approach we pursue is desirable since it is optimal in data reduction and conceptually simple, but is very computationally expensive. In terms of computations, in order to have $1^\circ$ pixel size, the entire sky consists of about $5\times10^4$ pixels and thus requires the inversion of a $5\times10^4$ by $5\times10^4$ matrix. While not computationally prohibitive, it is desirable to minimize the number of such inversions. Full-sky inversion has only becomes more relevant recently with the deployment of low-frequency radio interferometers with very short baselines (on the order of a few wavelengths), which are necessary to probe the structures on large angular scales. 

In this work, we propose a practical ``brute-force'' matrix-based method for imaging diffuse structure using interferometers. We pixelize the temperature distribution in the entire sky into a vector $\bm{s}$, and flatten all of the complex visibility data from all baselines and all times (as much as a whole sidereal day) into another vector $\bm{v}$. These two vectors are related by a set of linear equations described by a large matrix $\A$, determined by the primary beam and the fringe patterns of various baselines. We then deploy familiar tools for solving linear equations such as regularization and Wiener filtering to obtain a solution for the entire sky. This algorithm can potentially be applied to any interferometer, and it is especially advantageous for imaging diffuse structure rather than compact sources. The algorithm naturally synthesizes data from different times, without requiring any approximations such as a flat sky, a co-planar array, or a ``$w$-term''. It can therefore fill in ``missing modes'' that are otherwise undetermined using each snapshot individually. Furthermore, it allows for the optimal combination of data from different instruments in visibility space, rather than image space. Lastly, even for arrays with very high angular resolution for which the matrix inversion is impossible, this method can be used to calibrate its short baselines without needing an external sky model.

The rest of this paper is structured as follows. In \refsec{secdynamicmap}, we describe our method in detail. In \refsec{secsimulation}, we quantify the accuracy of the method using simulations for both MITEoR's maximally redundant layout and MWA's minimally redundant layout to provide intuitive understanding of its performance. In \refsec{secmap}, we apply the algorithm to our MITEoR data set to perform absolute calibration, and produce a  northern sky map with about $2^\circ$ resolution centered around 150\,MHz. We use the obtained map to compute spectral indices throughout our frequency range, as well as spectral indices when compared to 85\,MHz and 408\,MHz maps. To estimate the accuracy of our map, we compare it with both the 1970 Parkes map at 150\,MHz and the prediction of the GSM at 150\,MHz.

%%%%%%%%%%%%%%%%%%%%%%%%%%%%%
%%%%%%%%%%%%%%%%%%%%%%%%%%%%%
%%%%%%%%%%%%%%%%%%%%%%%%%%%%%
%%%%%%%%%%%%%%%%%%%%%%%%%%%%%
\section{Wide Field Interferometric Imaging}\label{secdynamicmap}
%%%%%%%%%%%%%%%%%%%%%%%%%%%%%
%%%%%%%%%%%%%%%%%%%%%%%%%%%%%
\subsection{Framework}\label{secframe}
We start with the interferometer equation for the visibility across the $i$th and $j$th antennas, following the conventions in \citet{FFTT}:
\begin{equation}
v_{ij} = \int{T_s(\bm{k})B(\bm{k})e^{i\bm{k}\cdot\bm{r}_{ij}}d\Omega},
\label{eqinter}
\end{equation}
where $T_s$ is the sky temperature, $B$ is the primary beam strength, $\bm{r}_{ij}$ is the baseline vector between the $i$th and $j$th antennae, and $\bm{k}$ is the wave vector of the electromagnetic radiation whose amplitude $|\bm{k}| = \frac{2\pi}{\lambda}$ and whose direction $\hat{\bm{k}}$ points towards the observer \citep{thompson_et_al2001}. In principle, one can generalize \refeq{eqinter} by including the full Stokes vector for the polarized sky temperature and a combination of Mueller matrices and Jones matrices for the primary beam, and follow the same algorithm described below to capture full polarization information. In this work, however, we limit our treatment to unpolarized sky temperature. We thus assume that the sky is unpolarized, and estimate its intensity using $xx$ and $yy$ polarization visibilities, which have different beam patterns. As such, there is potential polarization leakage from Stokes $Q$ and $U$ into $I$, but our wide primary beam and long observation time should limit such leakage.

Radio interferometry typically takes advantage of this equation by first performing a coordinate transformation on \refeq{eqinter} from $\bm{k}$ on the spherical surface to its projection on the $xy$-plane:
\begin{equation}\label{eqinter2}
v_{ij} = v(\bm{u}_{ij}) = \int{\frac{T_s(\bm{q})B(\bm{q})}{\sqrt{1-|\bm{q}|^2}}e^{i2\pi\bm{q}\cdot\bm{u}_{ij}}d\bm{q}},
\end{equation}
where $\bm{q}=\frac{\lambda}{2\pi}(k_x, k_y)$, and $\bm{u}_{ij} = \frac{\bm{r}_{ij}}{\lambda}$ \citep{tegmark_zaldarriaga2009}. By treating the visibilities, $v_{ij}$, and the beam-weighted sky image, $\frac{T_s(\bm{q})B(\bm{q})}{\sqrt{1-|\bm{q}|^2}}$, as a Fourier pair and performing 2D Fourier transforms on measured $v_{ij}$'s, radio interferometers have obtained high quality images of the sky with tremendous success. However, this Fourier approach comes with one important limitation. Generally speaking, the beam-weighted sky image is band-limited to the unit circle $|\bm{q}|\leq1$, so by the Nyquist theorem, one has to have $uv$ sampling finer than half a wavelength to avoid aliasing in the image. In reality, it is difficult to have any baselines shorter than half a wavelength due to the physical size of the antennas. What is more, since the shortest baseline has to be longer than the diameter of the antenna, the largest angular scale available is always smaller than the primary beam width, making aliasing inevitable. Mature techniques such as anti-aliasing filters \citep{aafilterbook} have been developed to solve this problem, but they are typically tailored for resolving compact structures rather than a diffuse background. The traditional $uv$-plane approach is therefore not ideal for imaging larger angular scales than the primary beam width.

To overcome this challenge, we follow \citet{JoshMM} and attack \refeq{eqinter} from a different angle. By discretizing the integral over sky angles into a sum over sky pixels indexed by $n$, and including measured visibilities from all times (such as 8 hours during a night's drift scan), \refeq{eqinter} becomes
\begin{equation}\label{eqinterdis}
v_{ijt} = \sum_{n}{T_s(\bm{k}_n)B(\bm{k}_n,t)e^{i\bm{k}_n\cdot\bm{r}_{ij}(t)}\Delta\Omega},
\end{equation}
where $\Delta\Omega$ is the pixel angular size, and we express all quantities in equatorial coordinates. Here the sky, $T_s$, is static, and $B(\bm{k}_i,t)$ and $\bm{r}_{ij}(t)$ change due to Earth's rotation and possibly the instrument's re-pointing such as in MWA's case. We then flatten visibilities from all times and baselines into one vector $\bm{v}$, flatten $T_s(\bm{k}_n)$ for all discretized directions in the sky into $\bm{s}$, and package the rotating beam and baseline information into a big matrix $\A$. \refeq{eqinterdis} now takes the form of a system of linear equations
\begin{equation}\label{eqvAs}
\bm{v} = \A\bm{s} + \bm{n},
\end{equation}
where we have now taken into account visibility noise $\mathbf{n}$ with mean zero and covariance $\langle\bm{n}\bm{n}^t\rangle \equiv \N$. Since the sky temperature is always real whereas $\A$ and $\bm{v}$ are complex, we ``realize'' the system by appending the imaginary part of $\A$ and $\bm{v}$ after their real parts, and double the noise variance. 

To optimally estimate $\bm{s}$, we use the minimum variance estimator \citep{tegmark1997b}
\begin{equation}\label{eqatniai}
\hat{\bm{s}} = (\A^t \N^{-1}\A)^{-1}\A^t \N^{-1}\bm{v},
\end{equation}
for which the error covariance matrix is
\begin{equation}\label{eqerrorcov}
\mathbf{\Sigma} \equiv \langle(\hat{\bm{s}} - \bm{s})(\hat{\bm{s}} - \bm{s})^t\rangle = (\A^t \N^{-1}\A)^{-1}.
\end{equation}
While this is an elegant set-up with a straightforward solution, there are two major technical difficulties with this approach: the size of $\A$ and the invertibility of $\A^t \N^{-1}\A$. We address these topics in detail in the next two sections.

%%%%%%%%%%%%%%%%%%%%%%%%%%%%%
%%%%%%%%%%%%%%%%%%%%%%%%%%%%%
\subsection{Constructing the $\A$-Matrix}\label{secAmatrix}
To obtain an intuitive understanding of the linear system described by \refeq{eqvAs}, we express the $n$th column of $\A$ according to \refeq{eqinterdis}, and change the coordinate system back to a rotating sky with a static beam and baselines:
\begin{align}
a_{ijt}^{(n)} =& B(\bm{k}_n,t)e^{i\bm{k}_n\cdot\bm{r}_{ij}(t)}\Delta\Omega\nonumber\\
=& B(\bm{k}_n)e^{i\bm{k}_n(t)\cdot\bm{r}_{ij}}\Delta\Omega.\label{eqacolumn}
\end{align}

For a given baseline pair $ij$, $a_{ijt}^{(n)}$ is the visibility over time on this baseline produced by the $n$th pixel in the sky if it had unit temperature. Stacking all baseline pairs together, the $n$th column of $\A$ is simply the set of visibilities the instrument would have measured if the sky only consisted of a single $n$th pixel of unit flux, with $0$ in all other pixels. Because the visibilities we measure are a linear superposition of contributions from all directions in the sky \citep{JoshMM}, finding an optimal solution to \refeq{eqvAs} is simply asking how much flux is needed in each pixel of the sky in order to jointly produce the visibilities we actually measured.

With this intuition in mind, we turn to the topic of how to pixelize the sky. Since $\A$ is not sparse, computing \refeq{eqatniai} will inevitably involve some form of inversion of $\A^t \N^{-1}\A$, whose computational cost scales as $n_{\text{pix}}^3$. We would therefore like the number of pixels to be as small as possible. On the other hand, in order to preserve accuracy in discretizing \refeq{eqinter} to \refeq{eqinterdis}, $\Delta\Omega$ needs to be smaller than square of the angular resolution of the instrument, where the angular resolution is roughly $\frac{\lambda}{d_{\text{max}}}$, and $d_{\text{max}}$ is the longest baseline length. When we pixelize the sky using the HEALPIX convention \citep{HEALPIX}, we thus need $n_{\text{side}}> \frac{d_{\text{max}}}{\lambda}$. For MITEoR, whose longest baseline is only about 15 wavelengths, we choose $n_{\text{side}}=64$, which translates to about $4.8\times10^4$ pixels, which is about the largest size that can be comfortably manipulated on a personal computer. To obtain higher resolution maps for other instruments with even longer baselines, one may choose to use non-uniform pixelization, which we discuss in \refapp{appdynamic}. It is worth pointing out that resolution alone does not decide the number of pixels needed; rather, what matters is the ratio of the primary beam width to the pixel width. Thus, the pixel number may not be very large if one wishes to image a small patch of sky at higher resolution with a narrow-beam instrument, depending on how quickly the sidelobes of the beam fall off away from its center.

%%%%%%%%%%%%%%%%%%%%%%%%%%%%%
%%%%%%%%%%%%%%%%%%%%%%%%%%%%%
\subsection{Regularization and Point Spread Functions}\label{secpsf}
For interferometric data sets, $\A^t \N^{-1}\A$ can be poorly conditioned and numerically non-invertible because the instrument is insensitive to certain linear combinations of the sky, so we insert a regularization matrix $\R$ (which we assume to be symmetric throughout this work) into \refeq{eqatniai}:
\begin{equation}\label{eqatnisiai}
\hat{\bm{s}}^\R = (\A^t \N^{-1}\A + \R)^{-1}\A^t \N^{-1}\bm{v}.
\end{equation}
We then substitute \refeq{eqvAs} into \refeq{eqatnisiai}, which leads to 
\begin{align}\label{eqpsf}
\hat{\bm{s}}^\R =& (\A^t \N^{-1}\A + \R)^{-1}\A^t \N^{-1}(\A\bm{s} + \bm{n})\nonumber\\
=& \mathbf{P}\bm{s} + (\A^t \N^{-1}\A + \R)^{-1}\A^t \N^{-1}\bm{n},\\
\langle\hat{\bm{s}}^\R\rangle =& \mathbf{P}\bm{s}
\end{align}
where we define the point spread matrix $\mathbf{P}\equiv(\A^t \N^{-1}\A + \R)^{-1}\A^t \N^{-1}\A$. The matrix $\mathbf{P}$ is not the identity matrix due to the insertion of $\R$, so each column of $\mathbf{P}$ acts as a point spread function (PSF) for the corresponding pixel in the true sky map $\bm{s}$. The introduction of regularization thus represents a comprise between the goal of completely removing the PSF and producing a map with favorable noise properties. The result is, in the language of traditional radio astronomy, a position-dependent synthesized beam. For each of the PSF, we can calculate the effective full width half maximum (FWHM) using 
\begin{equation}\label{eqfwhm}
\theta_\text{FWHM} = \sqrt{\pi^{-1}n_{\frac{1}{2}}\Delta\Omega},
\end{equation}
where $n_{\frac{1}{2}}$ is the number of pixels in the PSF whose absolute value is above half of the maximum pixel, and $\Omega_{n_\text{side}}$ is the angular area for each pixel. We use $\theta_\text{FWHM}$ to represent angular resolution throughout this work.

It is also worth noting that the sum of each row of $\mathbf{P}$ is a damping factor for the solution. For a fictitious uniform sky, if each row of $\mathbf{P}$ does not sum to unity, then the resulting solution will not retain the same uniform amplitude as the original fictitious sky. Thus, it is desirable to renormalize each row of $\mathbf{P}$ to sum to unity.

With the introduction of $\R$, the error covariance is now
\begin{align}\label{eqerrorcovR}
\mathbf{\Sigma}^\R =& \langle(\hat{\bm{s}}^\R - \mathbf{P}\bm{s})(\hat{\bm{s}}^\R - \mathbf{P}\bm{s})^t\rangle\nonumber\\
=& (\A^t \N^{-1}\A + \R)^{-1}(\A^t \N^{-1}\A)(\A^t \N^{-1}\A + \R)^{-1}\nonumber\\
=& \mathbf{P}(\A^t \N^{-1}\A + \R)^{-1}.
\end{align}
To quantify how well the predicted error in \refeq{eqerrorcovR} accounts for actual noise we obtain in either our simulations or maps using real data, we define the normalized error $\delta$ at the $i$th pixel as
\begin{equation}\label{eqchi}
\delta_i = \frac{|\hat{s}_i^\R-(\mathbf{P}\mathbf{s})_i|}{\sqrt{\Sigma_{ii}}}.
\end{equation}
Intuitively, $\delta$ represents the discrepancy between the recovered map and the ground truth in units of the expected noise level, and is thus expected to center around 1 in absence of any systematic effects. 

Now we discuss the choice of $\R$. Inserting $\R$ is equivalent to having prior knowledge of the sky, where the Bayesian prior for $\mathbf{s}$ has mean $\langle\mathbf{s}\rangle=\bm{s}^p=0$ and covariance matrix $\langle(\bm{s} - \bm{s}^p)(\bm{s}-\bm{s}^p)^{t}\rangle=\R^{-1}.$ If one has a sky model with well-characterized error properties (not the case for this work), there is a natural choice of $\R$ which we discuss in the next section. In the absence of such a model, however, the choice of $\R$ depends on the array layout, noise properties of the visibilities, as well as the trade-off between angular resolution and noise (we will demonstrate these qin much more detail through simulation in \refsec{secsimulation}). The simplest form of $\R$ is a uniform regularization matrix proportional to the identity: $\R = \epsilon^2\I$. $\epsilon^{-1}$ has the same units as $\bm{s}$, and can be compared to the noise level in the map solution: smaller $\epsilon^{-1}$ suppresses noisy modes more strongly since it will dominate the diagonal of $\mathbf{\Sigma}^\R$, but it also hurts angular resolution as it introduces wider point spread functions by bringing $\mathbf{P}$ farther away from the identity matrix. Therefore, $\epsilon$ is a tunable parameter deciding the trade-off between resolution and noise. In our simulations in \refsec{secsimulation} and map making using MITEoR data in \refsec{secmap}, we show various choices of $\epsilon$, and we leave investigations of the optimal choice of $\R$ for future work.

%%%%%%%%%%%%%%%%%%%%%%%%%%%%%
%%%%%%%%%%%%%%%%%%%%%%%%%%%%%
\subsection{Wiener Filtering and Incorporating Prior Knowledge}
If we have prior knowledge of the sky such as previous measurements, we would like to optimally combine the existing sky map $\bm{s}^p$, with our visibility measurements (this is not carried in our simulations or our final map product). We accomplish this by shifting our focus from $\bm{s}$ to $\bm{s}-\bm{s}^p$, and \refeq{eqatniai} becomes
\begin{equation}\label{eqatniaidiff}
\hat{\bm{s}} - \bm{s}^p= (\A^t \N^{-1}\A)^{-1}\A^t \N^{-1}(\bm{v} - \A\bm{s}^p).
\end{equation}
We then quantify the uncertainty in our prior knowledge, by defining the covariance matrix $\mathbf{S}\equiv\langle(\bm{s} - \bm{s}^p)(\bm{s}-\bm{s}^p)^{t}\rangle$, and we use $\R = \mathbf{S}^{-1}$ as the regularization matrix:
\begin{equation}\label{eqatnisiaidiff}
\hat{\bm{s}}^\R - \bm{s}^p= (\A^t \N^{-1}\A +  \mathbf{S}^{-1})^{-1}\A^t \N^{-1}(\bm{v} - \A\bm{s}^p).
\end{equation}
Our regularized estimate for $\bm{s}$ then becomes
\begin{equation}\label{eqatnisiaidiff2}
\hat{\bm{s}}^\R = (\A^t \N^{-1}\A +  \mathbf{S}^{-1})^{-1}\A^t \N^{-1}(\bm{v} - \A\bm{s}^p) + \bm{s}^p.
\end{equation}

There are three ways of understanding our choice of regularization $\R = \mathbf{S}^{-1}$. First, using the identity
\begin{equation}\label{eqXY}
(\X^{-1} + \Y^{-1})^{-1} = \X(\X + \Y)^{-1}\Y
\end{equation}
for invertible\footnote{We have mentioned that $\A^t\N^{-1}\A$ is usually numerically uninvertible, but as long as one excludes pixels never above the horizon, it should be formally invertible, with some eigenvalues very close to zero.} square matrices $\X$ and $\Y$, one can show that \refeq{eqatnisiaidiff} is equivalent to applying a Wiener filter $\W$ to \refeq{eqatniaidiff}:
\begin{equation}\label{eqatniaidiffW}
\hat{\bm{s}}^\R - \bm{s}^p= \W(\hat{\bm{s}} - \bm{s}^p),
\end{equation}
where  $\W = \mathbf{S}(\mathbf{\Sigma} + \mathbf{S})^{-1}$, with $\mathbf{\Sigma}$ defined in \refeq{eqerrorcov}.

Secondly, by using \refeq{eqXY}, one can show that \refeq{eqatnisiaidiff2} is equivalent to an inverse variance weighted average of the unregularized $\hat{\bm{s}}$ and the prior knowledge $\bm{s}^p$:
\begin{equation}\label{eqatnisiaidiffvariance}
\hat{\bm{s}}^\R = \mathbf{S}(\mathbf{\Sigma} + \mathbf{S})^{-1}\hat{\bm{s}} + \mathbf{\Sigma}(\mathbf{\Sigma} + \mathbf{S})^{-1}\bm{s}^p,
\end{equation}
where $\mathbf{\Sigma}$ and $\mathbf{S}$ are the noise covariance matrices for the unregularized $\hat{\bm{s}}$ and the prior knowledge $\bm{s}^p$, correspondingly. It is reassuring to see from \refeq{eqatnisiaidiffvariance} that $\hat{\bm{s}}^\R\to\bm{s}^p$ when $\mathbf{\Sigma} \gg \mathbf{S}$, and vice versa.

Lastly, \refeq{eqatnisiaidiff2} is equivalent to combining visibility data and a previous sky map through appending $\bm{s}^p$ to $\bm{v}$ and the identity matrix $\I$ to $\A$ in \refeq{eqvAs}, and solving it using \refeq{eqatniai} without any regularization.

Incorporating prior knowledge, especially in the form of existing pixel maps, can indeed fill in the ``missing modes'' and complement visibility data sets very well. However, we decided to leave demonstration of this part to future work for two reasons. First, there is no existing pixel map at the MITEoR frequency that covers the entire sky region observed by MITEoR. Second, our goal in this work is to demonstrate the effectiveness of our algorithm without access to additional data, despite the ``missing mode'' problem, so we therefore refrain from resorting to pixel maps to solve the invertibility challenges. Since we have made the data public, we very much hope that other authors will perform such improved regularization in the future. 

%%%%%%%%%%%%%%%%%%%%%%%%%%%%%
%%%%%%%%%%%%%%%%%%%%%%%%%%%%%
\subsection{Further Generalization}\label{secgeneral}
Throughout this section, we have, for clarity, limited our discussion to data sets in the form of baseline by time at a given frequency from one instrument. We generalize this to incorporate data sets from multiple frequencies and even multiple instruments. To synthesize multiple instruments at the same frequency, we simply append all the flattened data vectors together, and stack their corresponding $\A$ matrices in the same order. We then solve for the sky in one step using \refeq{eqatniai} or \refeq{eqatnisiai}. We demonstrate this through simulation in \refsec{secsimsummary}.

To synthesize multiple frequencies, we assume that the sky map only differs by an overall scaling factor throughout a given frequency range, and the accuracy approximation depends on how wide the frequency range is. If at each frequency $\nu$ we have 
\begin{equation}
\bm{v}_\nu = \A_\nu\bm{s}_\nu
\end{equation}
together with 
\begin{equation}
\bm{s}_\nu = f(\nu)\bm{s},
\end{equation}
we obtain
\begin{equation}
\bm{v}_\nu = f(\nu)\A_\nu\bm{s} = \A'_\nu\bm{s}.
\end{equation}

If we know $f(\nu)$, we can simply stack $\bm{v}_\nu$'s and $\A'_\nu$'s and solve for $\bm{s}$. In reality, although we do not know $f(\nu)$ very accurately, we can iterate this, where we start with an estimated $f_0(\nu)$ and at each iteration reevaluate $f(\nu)$ given by the best-fit ratio between $\bm{v}_\nu$ and $\A_\nu\bm{s}$ from the previous iteration. As a result, we can empirically obtain $f(\nu)$ in addition to $\bm{s}$. We demonstrate this in detail in \refsec{secmap}.

%%%%%%%%%%%%%%%%%%%%%%%%%%%%%
%%%%%%%%%%%%%%%%%%%%%%%%%%%%%
\subsection{Computational Discussion}\label{seccomputational}
As mentioned above in Sections~\ref{secAmatrix} and \ref{secpsf}, carrying out the matrix operations in \refeq{eqatnisiai} is computationally intensive. We therefore provide some computational details here for those readers interested in carrying out such calculations. 

First of all, it is important to judiciously choose the order of computations when following \refeq{eqatnisiai}, since a matrix-vector multiplication is much faster than a matrix-matrix multiplication. For example, it is much faster to calculate $\A^t(\N^{-1}\bm{v})$ than $(\A^t\N^{-1})\bm{v}$. 

Matrix multiplication is the dominant computational cost in the entire algorithm, whose cost scales as $n_\text{pix}^2n_\text{data}$, where $n_\text{data}$ is the number of data points, or the number of rows in the $\A$-matrix. Since typically $n_\text{data} > n_\text{pix}$, this is more expensive than inverting $(\A^t \N^{-1}\A + \R)$, which scales as $n_\text{pix}^3$. Due to the fact that the $\A$-matrix is neither sparse nor symmetric, iterative methods for linear system of equations, such as conjugate gradient, do not provide significant computational benefit. As a result, direct matrix multiplication of $\A^t\N^{-1}\A$ is preferable. Fortunately, matrix multiplication is trivially parallelizable, so parallelizing the matrix multiplication on multiple machines can significantly reduce computation time. The entire computation takes only of order a day on a modern linux computer with sufficient memory to hold the relevant matrices.

In addition, in order to compute the error covariance matrix defined by \refeq{eqerrorcovR}, direct inversion of $(\A^t \N^{-1}\A + \R)$ cannot be avoided. In our simulations discussed in the next section, we find that $(\A^t \N^{-1}\A + \R)$ has non-negligible off-diagonal terms, so approximating the matrix inversion using the inverse of the diagonal is not desirable. However, in circumstances when one does not need the error covariance matrix, such as when experimenting with the choice of $\R$-matrix, one can significantly reduce computational cost by taking advantage of the Cholesky decomposition, which provides a factor of 5-10 speed-up depending on the platform. 

Lastly, despite the goal of reducing computational cost, it is important to use double precision for matrix operations when implementing our method. Although neither the data nor the beam pattern are understood to any level near the numerical precision of single point floating numbers, and our simulation results do not rely on such precise knowledge (we calculate $\A$ using single precision, and the simulated data have noise), we found that using single precision during matrix operations has a significant and negative impact on the quality of our results because of the wide dynamic range of eigenvalues.

%%%%%%%%%%%%%%%%%%%%%%%%%%%%%
%%%%%%%%%%%%%%%%%%%%%%%%%%%%%
%%%%%%%%%%%%%%%%%%%%%%%%%%%%%
%%%%%%%%%%%%%%%%%%%%%%%%%%%%%
\section{Simulations}\label{secsimulation}
In this section, we perform simulations for both MITEoR and the MWA to demonstrate the algorithms we have described in the previous section. The distinct array layouts of MWA and MITEoR complement each other in demonstrating various aspects of our algorithm. The simulations are all based on just one night of observation on each instrument. To simulate visibilities, we pixelize the GSM using to the HEALPIX resolution $n_\text{side}=128$, but when we solve for the sky we use $n_\text{side}=32$ (a pixel size of $2^\circ$), so that we can quantify any potential errors introduced by having coarse pixels. The noise is simulated as Gaussian noise independent across baselines, frequencies and times, whose amplitude is set to the simulated autocorrelation over $\sqrt{\Delta \nu\Delta t}$, where $\Delta \nu$ is the bandwidth for each frequency bin, and $\Delta t$ is the integration time. For MITEoR, the noise variance is also scaled down by the redundancy factor (the number of antenna pairs sharing a baseline type) for each baseline type. We then follow Sections \ref{secframe}-\ref{secpsf} to compute maps and their error properties. 

%%%%%%%%%%%%%%%%%%%%%%%%%%%%%
%%%%%%%%%%%%%%%%%%%%%%%%%%%%%
\subsection{MITEoR Simulation}\label{secsimmiteor}
MITEoR is a highly redundant array. It consists of 64 dual-polarization antennas on a square grid with 3\,m spacing. For each polarization, it has 2,016 baselines with 112 unique baseline types. For this simulation, we include the shortest 102 out of its 112 unique baseline types, with baseline lengths between 3\,m and 25.5\,m. The primary beam model is numerically simulated as described in \citet{MITEoR}, and its FWHM is about $40^\circ$ near 150\,MHz. We simulate for a single night's observation in the local sidereal time (LST) range between 12 hours and 24 hours to resemble the LST coverage in \refsec{secmap}, at 150\,MHz, with 144 seconds integration, and 0.75\,MHz frequency bin width. The $\A$-matrix size for this set-up is $122,400\times9,725$. We choose $\epsilon^2\I$ as the regularization matrix, where $\epsilon^{-1} = 100$\,K (The result is not sensitive to the choice of $\epsilon$ within an order of magnitude, as shown later in \reffig{figepsilon}.).

The results we obtain are shown in the first column of \reffig{figsimulation}, in which the median noise is 14.9\,K. We computed the position-dependent FWHM using the PSF matrix, and found that 96\% of the pixels have FWHM less than the pixel size, which means that the resolution is limited by pixelization rather than the PSF. Note that Cygnus A (Cyg A) and Cassiopeia A (Cas A) cast noticeable ``ringing'' due to their extreme brightness and the effect of the PSF. Due to the coarseness of the pixelization in this section, we defer a demonstration of point source removal, as applied to Cyg A and Cas A, to \refsec{secmap}. Lastly, the normalized residual map shows that the errors in the recovered map are well described by the noise properties described by the noise covariance matrix $\mathbf{\Sigma}^\R$, with the exception of those regions with sharp features, such as pixels near Cyg A, Cas A, and the Galactic center. This is caused by the coarse pixelization, and can be remedied by decreasing the pixel size.

%%%%%%%%%%%%%%%%%%%%%%%%%%%%%
%%%%%%%%%%%%%%%%%%%%%%%%%%%%%
\subsection{MWA Simulation}\label{secsimmwa}
\begin{figure*}
	\centerline{\includegraphics[width=\textwidth]{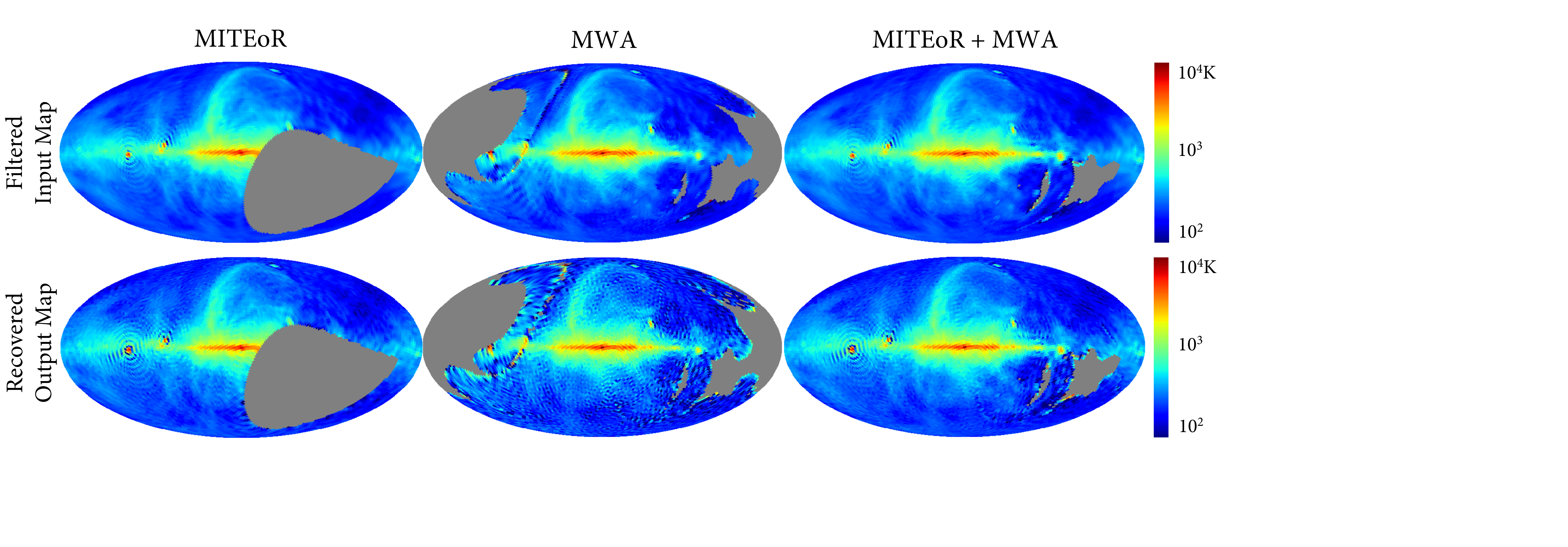}}
	\vskip-1.5mm
	\centerline{\includegraphics[width=\textwidth]{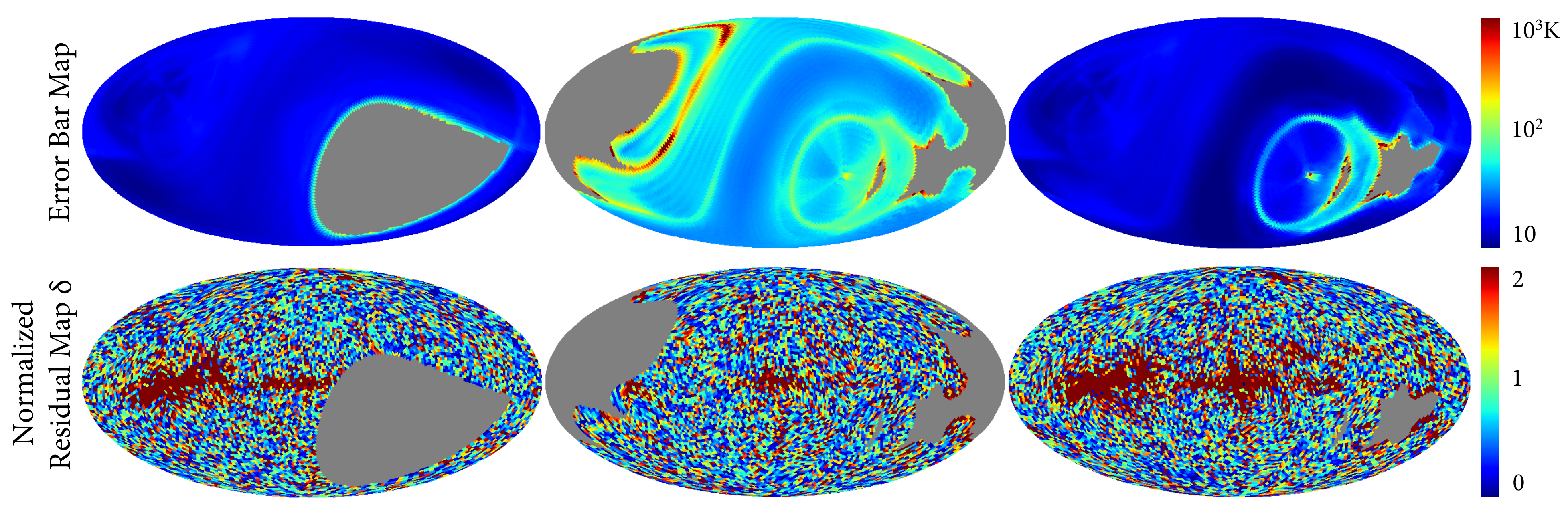}}
	\caption[Simulated results for MITEoR, the MWAcore, and the two combined.]{
		Simulated results for MITEoR (left), the MWAcore (middle), and the two combined (right). Filtered input GSM maps (top) are generated by applying the PSF matrix $\PSF$ to the GSM with $n_\text{side}=32$. Grey areas represent directions that are either never above the horizon or have negligible sensitivity. Recovered output maps are obtained by applying \refeq{eqatnisiai} to simulated noisy visibilities (assuming one night's observation and one frequency channel with $<1$\,MHz bandwidth), using uniform diagonal regularization matrices with $\epsilon^{-1}$ of 100\,K, 300\,K, and 100\,K respectively. The error bar maps are obtained by plotting the square roots of the diagonal entries in $\mathbf{\Sigma}^\R$, and the color scale is one order of magnitude smaller than the output maps. Lastly, the normalized residual $\delta$ maps represent the ratio of the actual error in our maps to the error bars, as defined in \refeq{eqchi}, and their values center around 1 as we expect. Here and in the subsequent figures, all maps are in Mollweide projection centered on the Galactic center.
		\label{figsimulation}
	}
\end{figure*}

MWA is a minimally redundant array, with 8,128 different baselines. For this simulation, we use the shortest 195 cross-correlation baselines, with baseline lengths between 7.7\,m and 25.5\,m (referred to as MWAcore from hereon). We limit the baseline lengths in order to use the same $n_\text{side}=32$ pixelization as the MITEoR case, as longer baselines requires a much finer pixelization. The primary beam model is obtained by calculating the analytic expression for phased array of short dipoles, and the FWHM is about $14^\circ$ around 150\,MHz. We simulate for a single night's observation in the LST range between 12 hours and 24 hours, at 150\,MHz, with 144 seconds integration, and 0.64\,MHz frequency bin width. The beam remains in the zenith-pointing drift scan mode for the entire observation, which is not how the MWA typically operates (but doable). The $\A$-matrix size for this set-up is $234,000 \times 9,785$. We choose $\epsilon^2\I$ as the regularization matrix, where $\epsilon^{-1} = 300$\,K. The results we obtain are shown in the second column of \reffig{figsimulation}. The median noise in the map is 47.8\,K, and the higher noise lines follow the trajectory of the nulls between the main lobe and the first side lobe, as well as between the first and second side lobes. In terms of the angular resolution, 89\% of the sky is limited by pixelization. 

%%%%%%%%%%%%%%%%%%%%%%%%%%%%%
%%%%%%%%%%%%%%%%%%%%%%%%%%%%%
\subsection{Simulation Discussion}\label{secsimsummary}
\begin{figure*}
	\centerline{\includegraphics[width=\textwidth]{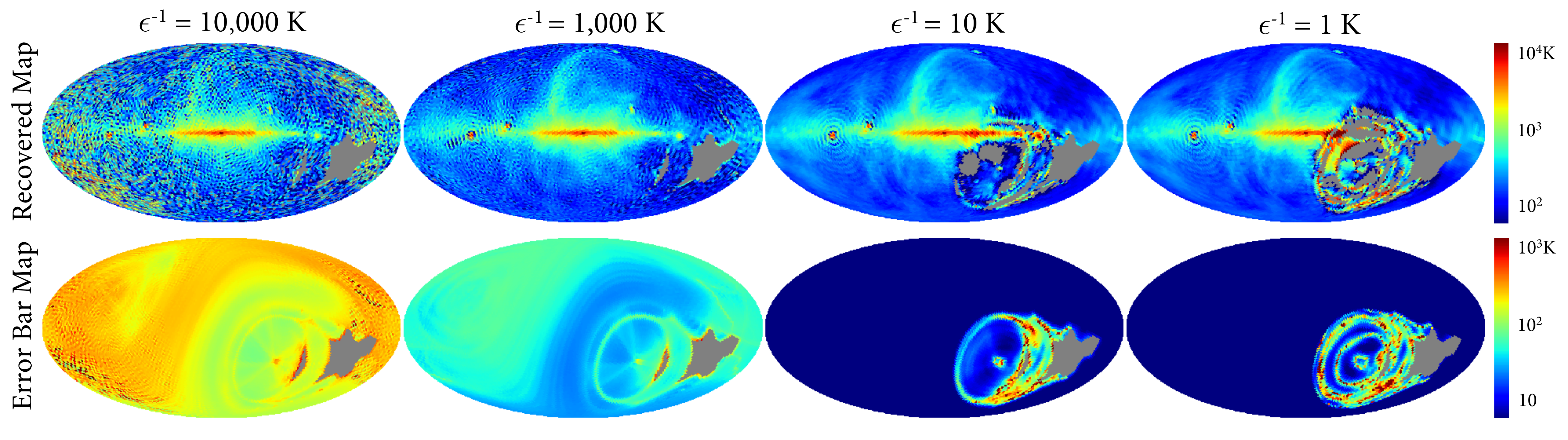}}
	\caption[Output maps recovered in simulation and their error bars using different regularization matrices.]{
		Output maps recovered in simulation and their error bars ($\sqrt{\mathbf{\Sigma}_{ii}^\R}$), using the same set of data but different regularization matrices. As $\epsilon$ increases over 5 orders of magnitude ($\epsilon^{-1}=100$\,K is shown in the third column of \reffig{figsimulation}), the noise is significantly suppressed, but the properties of PSF matrix become less desirable: we start to see negative temperatures near the null regions of the MWA beam. This is because stronger regularization suppresses the ``missing modes'' more heavily, and as the large scale structures near the MWA's side lobe region gets suppressed, more of that region becomes negative. In addition, strong regularization creates worse ``ringing'' near Cyg A and Cas A, and even near the north pole spur in the 1\,K case.
		\label{figepsilon}
	}
\end{figure*}

By combining the MWAcore visibilities and the MITEoR visibilities, and using a uniform regularization $\epsilon^{-1}=100$\,K, we obtain another map shown in the third column of \reffig{figsimulation}, whose median noise is 14.6\,K. In this case, 97\% of the pixels' resolution is limited by pixelization. To demonstrate the effect of tuning the parameter $\epsilon$ in the regularization matrix, we show a few different choices of $\epsilon$ in \reffig{figepsilon}. As the regularization varies from two orders of magnitude too weak to two orders of magnitude too strong, overall noise level decreases, but stronger PSFs start to overly suppress the less sensitive regions near the nulls of the MWA beam and make the temperature negative, and create worse ``ringing'' around Cyg A and Cas A.

As shown in \reffig{figsimulation}, with just one night's data, our algorithm can produce high quality diffuse structure maps using either the MWAcore or MITEoR. With the MWAcore's set-up where the shortest baseline is longer than 7 wavelengths, the algorithm successfully determines large scale modes such as the overall amplitude of the map. On the other end, the longest baselines we include are 13 wavelengths long, which naively translates to about $5^\circ$ resolution, but the matrix approach recovers angular scales smaller than $2^\circ$, as both simulations show that the resolution is limited by the $2^\circ$ pixel size. As we will show in \refsec{secmap} using $1^\circ$ pixels, the true resolving power of these baselines are between $1^\circ$ and $2^\circ$.

Comparing the noise maps of the MWAcore and MITEoR, where MITEoR's 14.9\,K median noise level is slightly lower than MWAcore's 47.8\,K, we see that the overall noise level is not very sensitive to the number of baseline types, the baselines' length distribution, or the primary beam shape. Although the collecting area of MWA tiles is 16 times larger than each bow-tie antenna used in MITEoR, it is offset by the fact that there are effectively about 2000 baselines used in the MITEoR simulation (with 112 unique baseline types) compared to the MWAcore's 195, so the noise levels are comparable in both cases. In contrast to the overall noise level, the spatial patterns of noise do depend heavily on the primary beam pattern, which is not surprising, since the primary beam pattern heavily influences the instrument's sensitivity to different parts of the sky.

The $\delta$ maps show that the error properties are well characterized by $\mathbf{\Sigma}^\R$, with the majority of the $\delta$ values less 2. We see that although the visibilities are simulated using the GSM with $n_\text{side}=128$, the crude pixelization of $n_\text{side}=32$ is not introducing significant errors, other than in regions near Cyg A, Cas A, or the Galactic center. The importance of pixelization errors depends on the baseline lengths included, as well as the amount of noise in the visibilities. We find that pixelization errors become significant if we increase the baseline length threshold by another 50\% to about 40\,m. We also expect the pixelization error to become dominant if the visibilities have much lower noise, such as when they are averaged over 100 nights as opposed to a single night used here.

Given the result that just one night's data from MWAcore's zenith pointing scan can determine more than half of the sky to better than 50\,K precision, MWA has the potential to make a high quality ($<10$\,K noise through multiple nights' observations) southern sky map with multiple beam pointings and multiple nights throughout the year to fill the sky. In addition, since MWAcore only includes a small fraction of MWA's baselines, a much higher angular resolution (less than a degree) is also achievable with finer pixelizations.

%%%%%%%%%%%%%%%%%%%%%%%%%%%%%
%%%%%%%%%%%%%%%%%%%%%%%%%%%%%
%%%%%%%%%%%%%%%%%%%%%%%%%%%%%
%%%%%%%%%%%%%%%%%%%%%%%%%%%%%
\section{New Sky Map}\label{secmap}
%%%%%%%%%%%%%%%%%%%%%%%%%%%%%
%%%%%%%%%%%%%%%%%%%%%%%%%%%%%
As we have demonstrated in the previous section, the new imaging method works well on simulated data from both MITEoR and MWAcore. In this section, we apply the method to real data collected by MITEoR, and produce a Northern sky map at 150\,MHz.

\subsection{MITEoR Instrument and Data Reduction}\label{subsecmiteordata}
\begin{figure*}
	\centerline{\includegraphics[width=\textwidth]{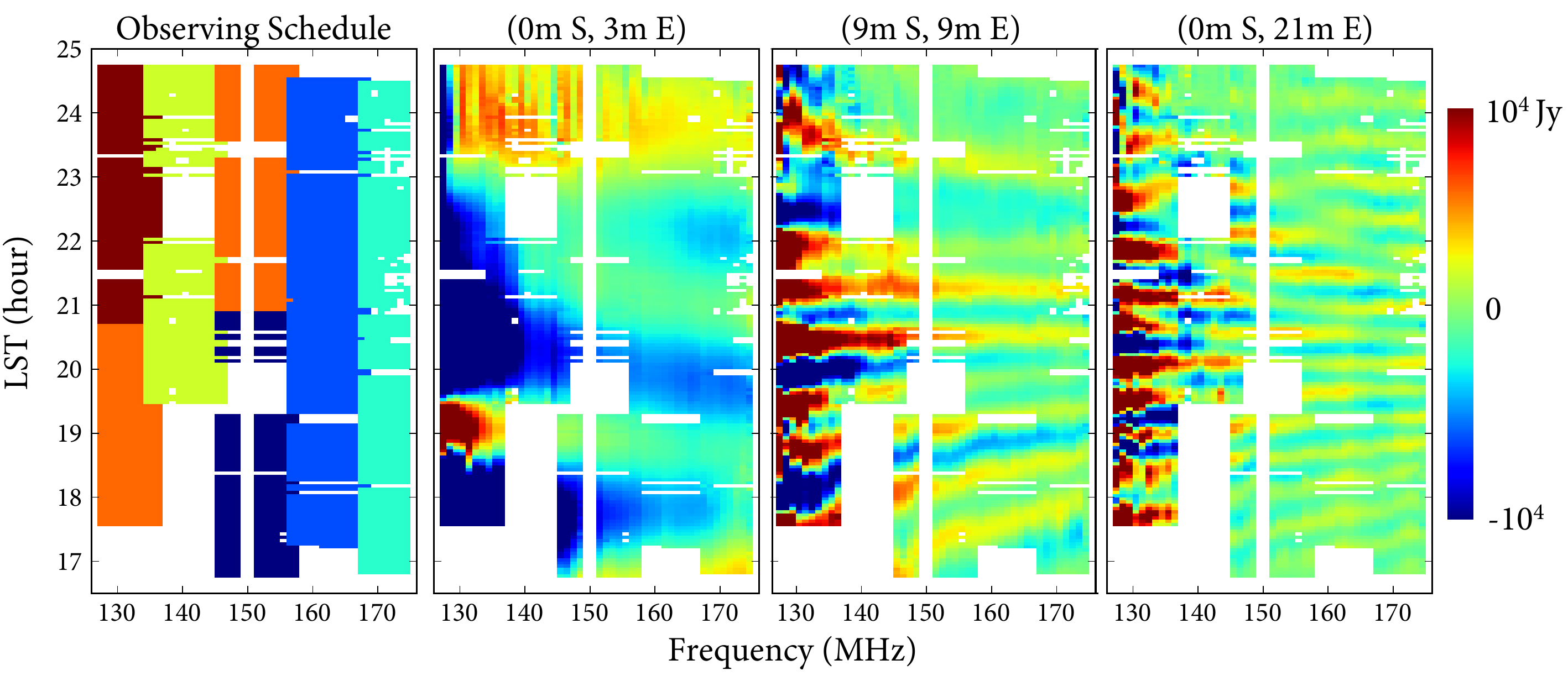}}
	\caption[MITEoR's observing schedule and a small subset of the MITEoR data product.]{
		MITEoR's observing schedule on the left and three sets of visibilities in the MITEoR data product on the right. Each of the six colors represent one of the nights between July 27th and August 2nd, 2013. Midnight corresponds to LST at roughly 21 hours. The real part of visibilities over time and frequency on three baselines of very different lengths are shown here. The white gaps in frequency and time are RFI events automatically flagged by the redundant calibration pipeline.
		\label{figmiteordata}
	}
\end{figure*}

As mentioned above, MITEoR \citep{MITEoR} is a compact radio interferometer with 64 dual-pol antennas, deployed in July 2013 in The Forks, Maine (latitude $45.3^\circ$). The antennas are identical to individual MWA bow-tie antennas (without beam forming as tiles), and the full width half maximum (FWHM) of the primary beam is about $40^\circ$ throughout our frequency range. The data used in this work are collected with an 8 by 8 square grid array configuration with 3\,m spacings. The 8 bit correlator cross-correlates all 128 antenna-polarizations (each bow-tie antenna has two polarizations as outputs), with integration time of 2.7 seconds, instantaneous bandwidth of 12.5\,MHz (tunable between 125 and 185 \,MHz), and frequency bin width of 50\,kHz. The data used in this work were collected through 7 observing sessions, as shown in the left panel of \reffig{figmiteordata}. 

We first perform redundant calibration on the raw data, using our redundant calibration pipeline described in \citet{MITEoR}, with further improvements described in \refapp{appdegen}. Redundant calibration compresses the data in the baseline direction from 2016 cross-correlation visibilities per snapshot to 112 unique baseline types, and automatically flags bad antennas, baselines, frequencies, and time stamps from the data. It is worth noting that redundant calibration uses only the self-consistency between redundant baselines, and does not use any sky model. After redundant calibration, we further compress the data to 0.75\,MHz frequency bin width by averaging every 15 frequency bins. We then average over the time direction in 2 minute intervals. We empirically estimate noise during the time averaging step by performing linear fitting over 2 minutes of data and calculating the residual power.

At this stage, we have a data cube of 4 polarizations by 75 frequency bins by 240 time steps by 112 baselines. Since we have not used any sky information, the data is not yet absolute-calibrated, meaning that for each of the $4\times75=300$ different polarization-frequency data sets, we have 3 undetermined numbers: one overall amplitude, and two re-phasing degrees of freedom. These numbers cannot be determined without performing absolute calibration (as opposed to redundant calibration) with a sky model, which we describe in the next two sections.

%%%%%%%%%%%%%%%%%%%%%%%%%%%%%
\subsubsection{Absolute Amplitude Calibration}\label{subsecampcal}

The goal of absolute calibration is to determine three numbers (an overall amplitude and two re-phasing degrees of freedom) for every time by baseline data set, where each data set consists of more than $10^4$ visibility measurements. This is a drastically overdetermined system given a sky model. In this section, we discuss how we determine the overall amplitude. Since we intend to use the map obtained in this work to improve the GSM in a future work, we choose not to use the GSM as our sky model. Rather, we use Cyg A and Cas A as our calibrators. After we obtain a map in \refsec{subsecfinalmap}, we will lock the amplitude of our map to the Parkes map at 150\,MHZ \citep{Landecker1970Parkes}, so the amplitude calibration and its error will only affect our spectral index results, not the amplitude of the map.

Our amplitude calibration is based on extrapolating the frequency-dependent secular decreasing flux models of Cas A \citep{Viny14} and the spectrum of Cyg A from \citet{Viny06}. Cyg A is an ultra-luminous, jet-powered, radio-loud galaxy. For our Cyg A calibration spectrum, we use the model in Eq.~6a in \citet{Viny06} of a transparent source, with a power law spectrum with spectral index $\alpha$, observed through the absorbing ionized gas of our Galaxy. The data point used in the model at 152\,MHz has a reported 3\% error in \citet{Parker68}, and propagating the model parameters' error bars leads to a maximum of 3.2\% error in our frequency range.

The frequency dependence of the decreasing flux of Cas A has been widely studied (see \citet{Helm09} and references, therein), and we adopt the empirical model in \citet{Viny14} that fits to the accumulated published data taken from 1961 to 2011 from about 12\,MHz\,-\,93\,GHz, including their most recent observations. The spectrum of Cas A is evaluated in frequency range 125-175 MHz, using the fitted function in Eqs.~15 and 16 in \citet{Viny14}. With this spectrum modeled at epoch 2015.5, and the model for the frequency dependent secular decrease in Eqs.~9 and 10 of \citet{Viny14},  we evaluate the spectrum for Cas A during MITEoR observations in August 2013, approximately two years earlier. The largest source of error in the radio spectrum model of Cas A comes from our lack of complete understanding of the behavior of this supernova remnant. In an evaluation of possible periodic deviations, \citet{Helm09} identify 4 possible modes ranging from 3 to 24 years, in a slightly lower frequency range of interest, $38-80$ MHz, contributing to flux deviations from a secular decrease of up to 15\%. 

We calibrate in the LST range between 19 and 23 hours, during which Cyg A's elevation ranges from $59^\circ$ to $85^\circ$, and Cas A from $36^\circ$ to $65^\circ$. To minimize error introduced by diffuse structures, we only include for calibration baselines longer than 8.6 wavelengths, which are the longest 9 baselines at the lowest frequency and 34 at the highest frequency. After fitting using Cas A and Cyg A, we find about 15\% residual on the visibilities. Since the errors introduced by the Galactic plane are averaged down over different baseline types, we estimate our amplitude calibration to go down by a factor equal to the square root of the number of baselines used. Thus, our absolute calibration has an overall error of about 5\% at the lowest frequency and 2.5\% at the highest frequency, relative to the calibrators. 

%%%%%%%%%%%%%%%%%%%%%%%%%%%%%
\subsubsection{Absolute Phase Calibration}\label{subsecphasecal}
As discussed in \citet{MITEoR}, the two re-phasing degrees of freedom (or re-phasing degeneracies) in the visibility space correspond to shifting the beam-weighted sky image $\frac{T_s(\bm{q})B(\bm{q})}{\sqrt{1-|\bm{q}|^2}}$, and cannot be determined using only the self consistency of visibility data without a sky model. However, this is only true for isolated snapshots in time. For instruments with large fields of view such as MITEoR, rotation of the sky does not exactly translate into shifting the beam-weighted sky image in the projected $\mathbf{q}$-plane, so a constant shift of the beam-weighted sky image caused by constant re-phasing cannot be consistent with a rotating sky. Therefore, we can use a procedure conceptually similar to self-cal to determine the re-phasing: we first image using the visibility without correcting for the re-phasing degrees of freedom, then use the image we obtain to solve for the re-phasing, and iterate until convergence. In theory, this can be done without any prior sky model, but since each iteration can be computationally expensive, we use the GSM to provide the initial re-phasing solution, and start iterating from there. It is worth noting that, at a given frequency and polarization, unlike self-cal which is solving for, say, 128 antenna calibration parameters at every time stamp, here we are only solving for 2 numbers for an entire observing session, so this iterative algorithm has negligible impact on the validity of \refeq{eqerrorcovR}. For more detailed discussions of redundant phase calibration, we refer the readers to \citet{MITEoR} and Appendix~\ref{appdegen} of this work.

In addition, even for instruments whose array layout prevents making usable images, this ``self-cal'' approach can be applied to remove re-phasing degeneracies. Rather than inserting regularization matrices to make $\A^t\N^{-1}\A$ invertible, if the goal is to calibrate out the re-phasing degeneracies, we can simply use a pseudo-inverse for $\A^t\N^{-1}\A$. For example, for a redundant array with no short baselines, a pseudo-inverse will remove large scale structures in the image, but this has no effect when the resulting image is used as a model to simulate visibilities on those long baselines, so ``self-cal'' should work just as well. 

%%%%%%%%%%%%%%%%%%%%%%%%%%%%%
\subsubsection{Cross-talk Removal}\label{subseccrosstalk}
We define cross-talk as additive offsets on visibilities that are proportional to the amplitude of auto-correlation, with a small but constant proportionality coefficient. In theory, the cross-talk terms can be solved for in an iterative fashion similar to how we determine the re-phasing degeneracies in the previous section. However, due to the level of thermal noise and systematics present in our data set, cross-talk on our shortest baselines is highly degenerate with having a bright stripe in the trajectory of our local zenith. Thus for this work, we use the GSM to perform cross-talk removal. We use the GSM to simulate all the visibilities we measure, and for each visibility time series, we solve for the best fit using the GSM model visibility and the auto-correlation, and subtract the auto-correlation component from our data. Thus, for each visibility time series over a few hours, we use the GSM to fit and remove one degree of freedom corresponding to cross-talk.

\begin{figure*}
	\centerline{
		\includegraphics[width=.45\textwidth]{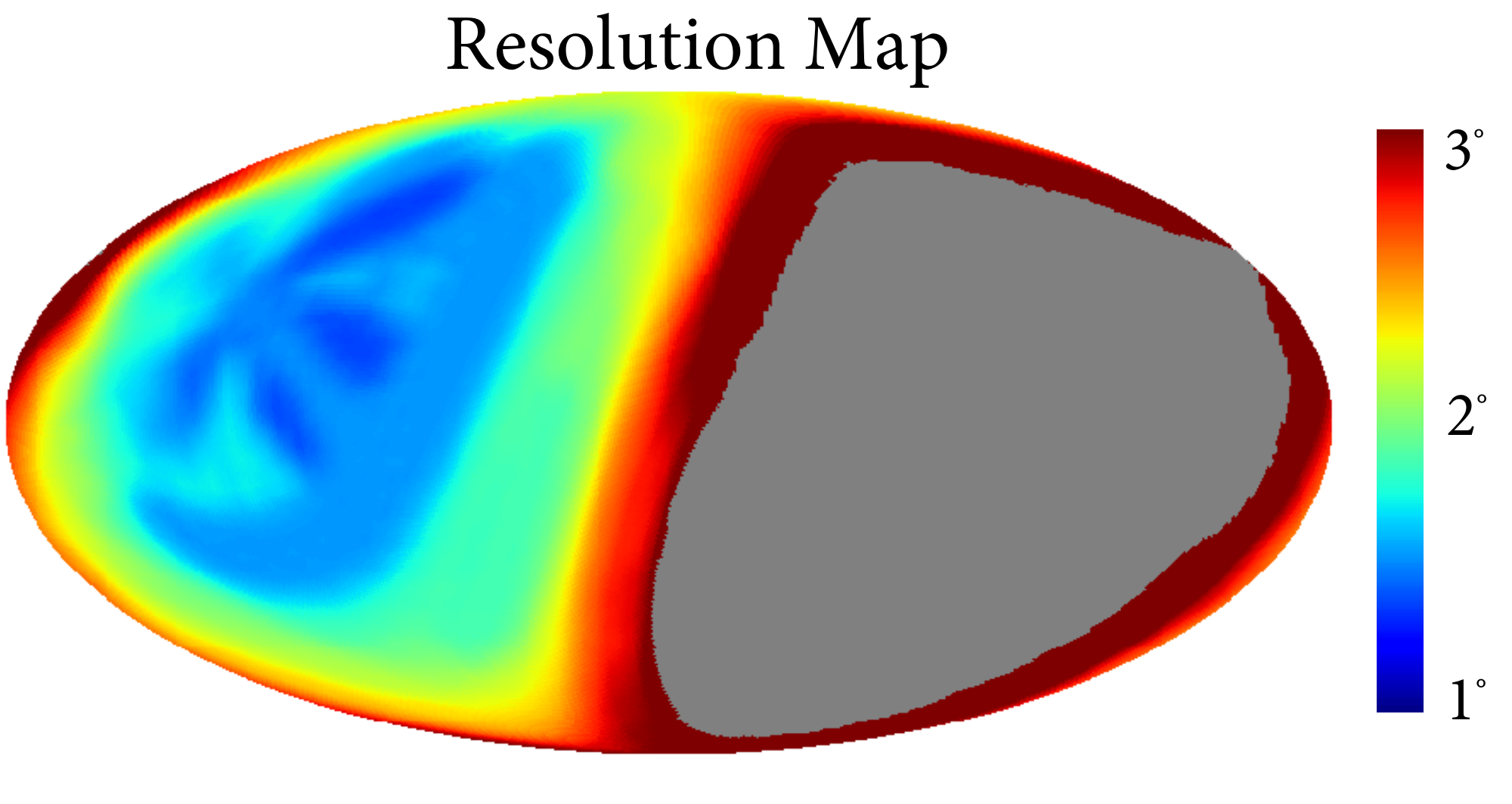}
		\includegraphics[width=.45\textwidth]{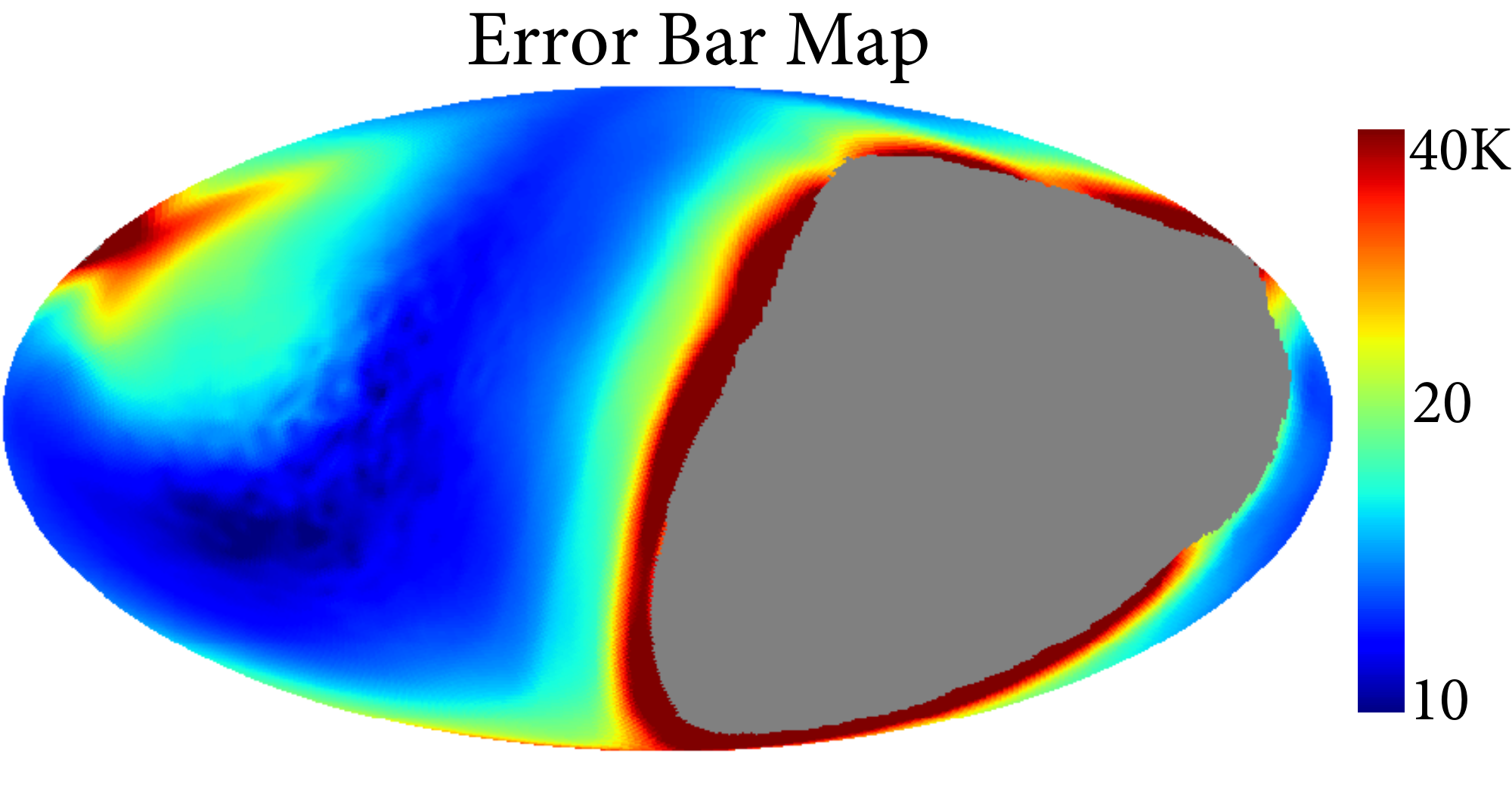}
	}
	\centerline{\includegraphics[width=\textwidth]{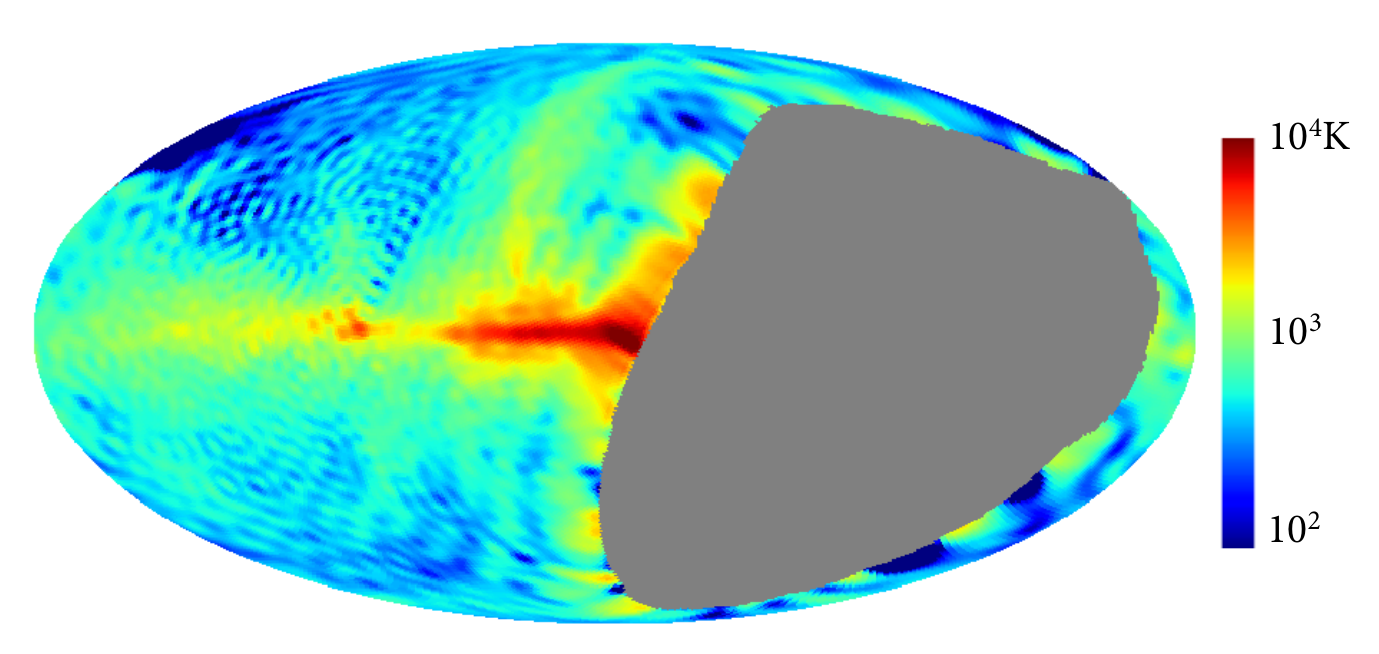}}
	\centerline{
		\includegraphics[width=.48\textwidth]{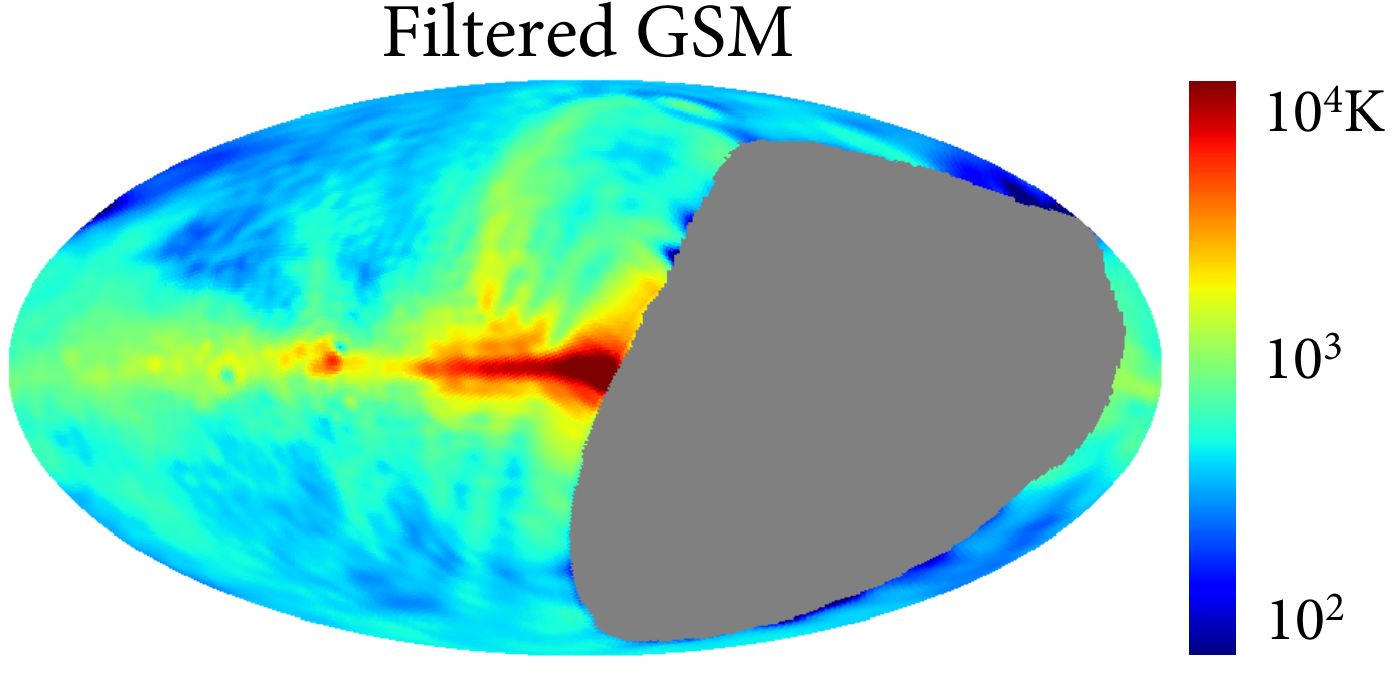}
		\includegraphics[width=.48\textwidth]{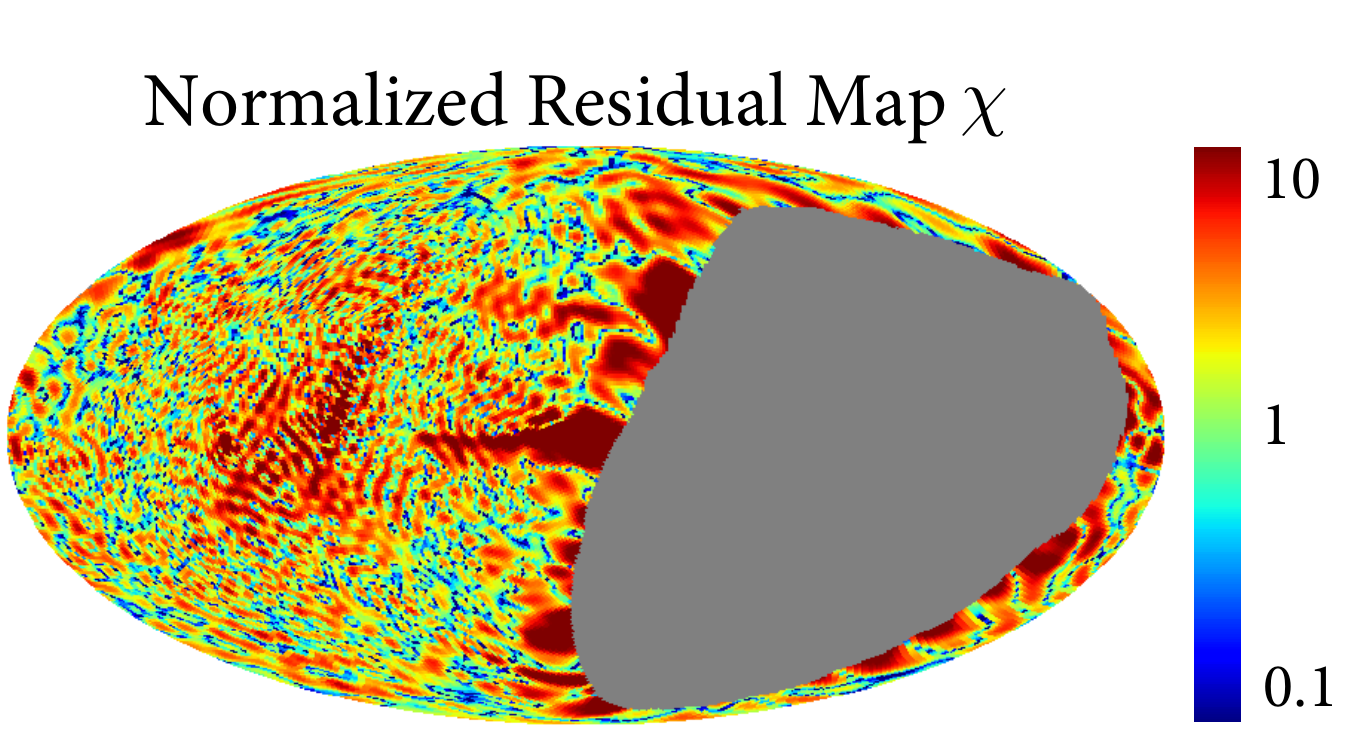}
	}
	\caption[MITEoR map result.]{
		Top left: angular resolution map obtained from FWHM of the point spread functions. Top right: error bar map obtained from $\sqrt{\Sigma_{ii}}$ based on empirically estimated noise covariance $\N$. Mid: the northern sky map at 150\,MHz, averaged from 128.5\,MHz to 174.5\,MHz, with Cas A and Cyg A removed using CLEAN. Bottom left: The GSM at 150\,MHz, with the PSF matrix applied. Bottom right: $\delta$ as defined in \refeq{eqchi}, which represents the ratio between the difference between the maps and our map's error bars. The median $\delta$ is 2.16.
		\label{figmapuber}
	}
\end{figure*}
%%%%%%%%%%%%%%%%%%%%%%%%%%%%%
%%%%%%%%%%%%%%%%%%%%%%%%%%%%%
\subsection{Northern Sky Map Combining Multiple MITEoR Frequencies}\label{subsecfinalmap}
We apply the algorithm described in \refsec{secdynamicmap} on the MITEoR data to obtain our Northern sky map. We have shown in \refsec{secsimulation} that the MITEoR data can in principle make high quality maps at individual frequency bins, but due to the systematics present in our data, which we discuss more in the next section, we are not able to make high quality maps using each individual frequency alone. Since in our frequency range the diffuse emission is dominantly synchrotron, which follows a smooth power law, we use techniques described in \refsec{secgeneral} to combine multiple frequencies to form a single map with an overall spectral index, as well as the beam-averaged spectral index as a function of time. We pixelize the sky to HEALPIX $n_\text{side}=64$. Since the size of the $\A$-matrix is proportional to the number of frequency bins, and including the entire data set amkes the size of our $\A$-matrix too large, we include only one out of every 5 frequency bins throughout the frequency range of 128.5\,MHz to 174.5\,MHz. This forms an $\A$-matrix of size roughly $6\times10^5$ by $4\times10^4$. Since $\A$ has an order of magnitude more rows than columns, computing $\A^t\N^{-1}\A$ is the speed bottleneck and takes more than 2 days when parallelized on a single CPU. In comparison, inverting $\A^t\N^{-1}\A$ takes about 3 hours.

In order to calculate the relative amplitude at different frequencies, to solve for the re-phasing degeneracies in the data, and to empirically estimate the level of noise and systematics in each data set, we iterate the process described in \refsec{secpsf} until the amplitudes converge to within 0.1\% and re-phasings within $0.1^\circ$. At each iteration, we calculate visibilities using the solution from the previous iteration as a model, and for each frequency we fit for the re-phasing, the relative amplitude, as well as the overall error RMS. Both the best fit amplitude and the error RMS are used to re-weigh the noise covariance matrices for each frequency. When iterating, we prioritize the map's accuracy in modeling visibilities over its noise properties, so we choose a weaker regularization of $\epsilon^{-1}=1500$~K. For the final map we choose a stronger regularization with $\epsilon^{-1}=300$~K to obtain lower noise in the map at a cost of lower resolution in the noisy areas. The map's overall amplitude is locked to the Parkes map at 150\,MHz \citep{Landecker1970Parkes} using the overlapping region. Cyg A, Cas A, and their ``ringing'' are removed using the CLEAN algorithm \citep{CLEAN}. The map we obtain together with its angular resolution and error bars are shown in \reffig{figmapuber}.%657740 41832

%%%%%%%%%%%%%%%%%%%%%%%%%%%%%
%%%%%%%%%%%%%%%%%%%%%%%%%%%%%
\subsection{Error Analysis}\label{subsecerror}
\reffig{figmapuber} shows that the map we obtain agree very well with the prediction of the GSM at 150\,MHz. In this section, we discuss the errors in our map and their possible causes in more detail. In terms of overall amplitude, the map's overall amplitude is locked to the Parkes 150\,MHz equatorial map \citep{Landecker1970Parkes}, which has 20\,K uncertainty in zero level and 4\% in temperature scale. In order to compare detailed structures in our result with the Parkes map and the GSM, we calculate the normalized error $\delta$ maps, shown in \reffig{figmapuber}. The median $\delta$ compared to the filtered GSM is 2.16, and 2.79 to the unfiltered Parkes map, which are slight higher than what one might expect. 

There are a few factors that make the median $\delta$ higher than 1. Firstly, our modeling of our instrument is not perfect, leading to error in our $\A$-matrix. As we investigated in detail in \citet{MITEoR}, our beam model has up to 10\% error in some directions. Our empirically estimated visibility errors are dominated by slowly varying modes, and beam mis-modeling is the most likely cause. Another cause of error is the averaging over frequency, which assumes constant spectral index over the sky. As we show in much more detail in the next section, the spectral index changes by as much as 0.5 from the Galactic plane to out-of-plane regions, so this introduces an error for the edge frequencies on the order of $(175\,\text{MHz}/150\,\text{MHz})^{0.5} - 1 = 8\%.$ Moreover, as we have seen in our simulation results in \reffig{figsimulation}, pixelization can also cause un-modeled error near high temperature regions such as the point sources (Cas A and Cyg A) and the Galactic center. Lastly, neither the GSM nor the Parkes map are the ground truth: the GSM at 150\,MHz is essentially an interpolation product between a 45\,MHz sky map and a 408\,MHz sky map, with an estimated relative error of 10\% \citep{GSM}. For the equatorial Parkes map, due to its $2.2^\circ$ resolution, we cannot apply the PSF matrix to it before comparing it to our result. The lack of PSF on the Parkes map leads to higher error in the comparison, which we think is what makes the median $\delta$ higher for the unfiltered Parkes map than for the GSM. 

%%%%%%%%%%%%%%%%%%%%%%%%%%%%%
%%%%%%%%%%%%%%%%%%%%%%%%%%%%%
\subsection{Spectral Index Results}\label{subsecspectral}
%------------------------------------------------------------
\subsubsection{Spectral Indices from 128\,MHz to 175\,MHz}
\begin{figure}
	\centerline{\includegraphics[width=.45\textwidth]{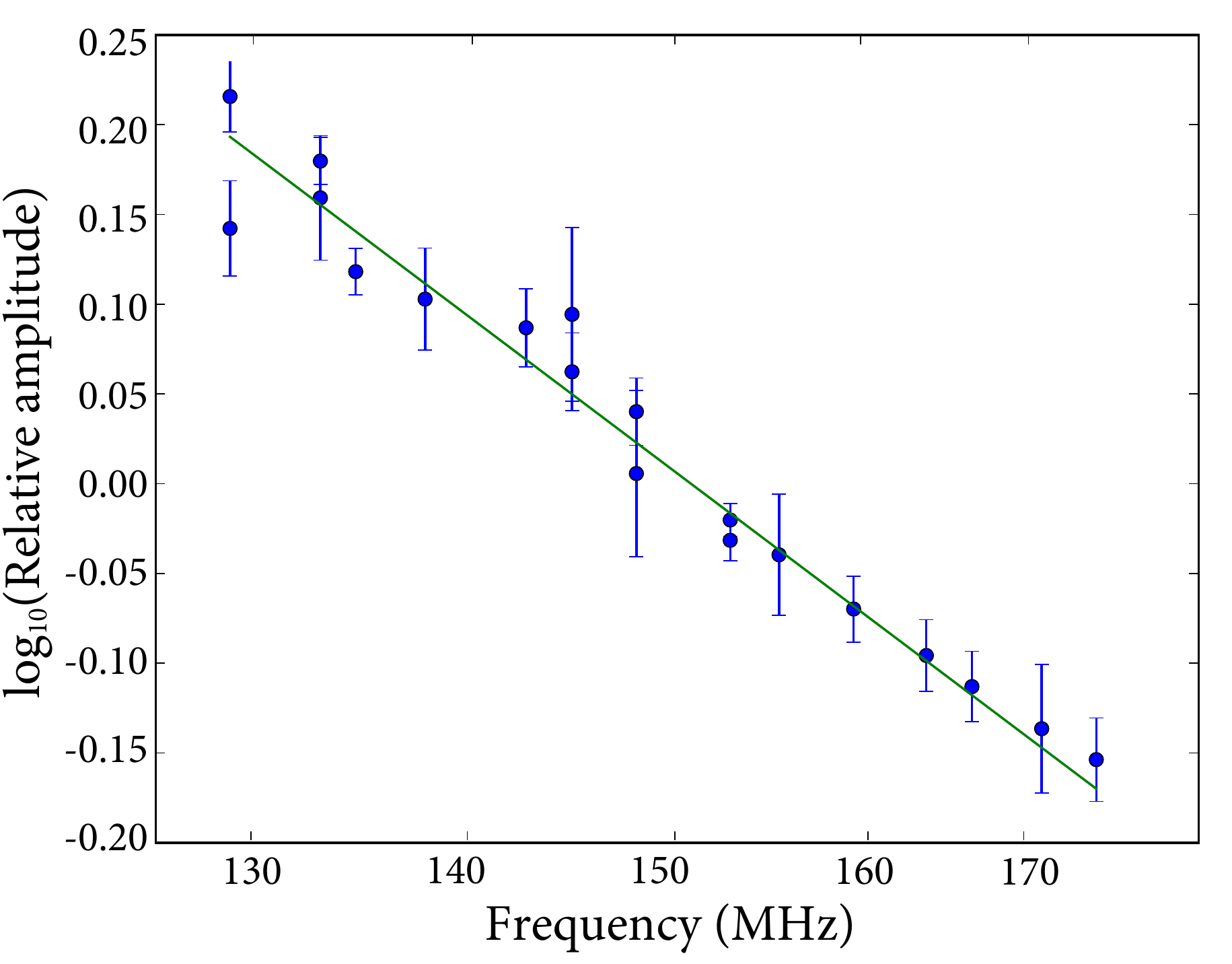}}
	\centerline{\includegraphics[width=.45\textwidth]{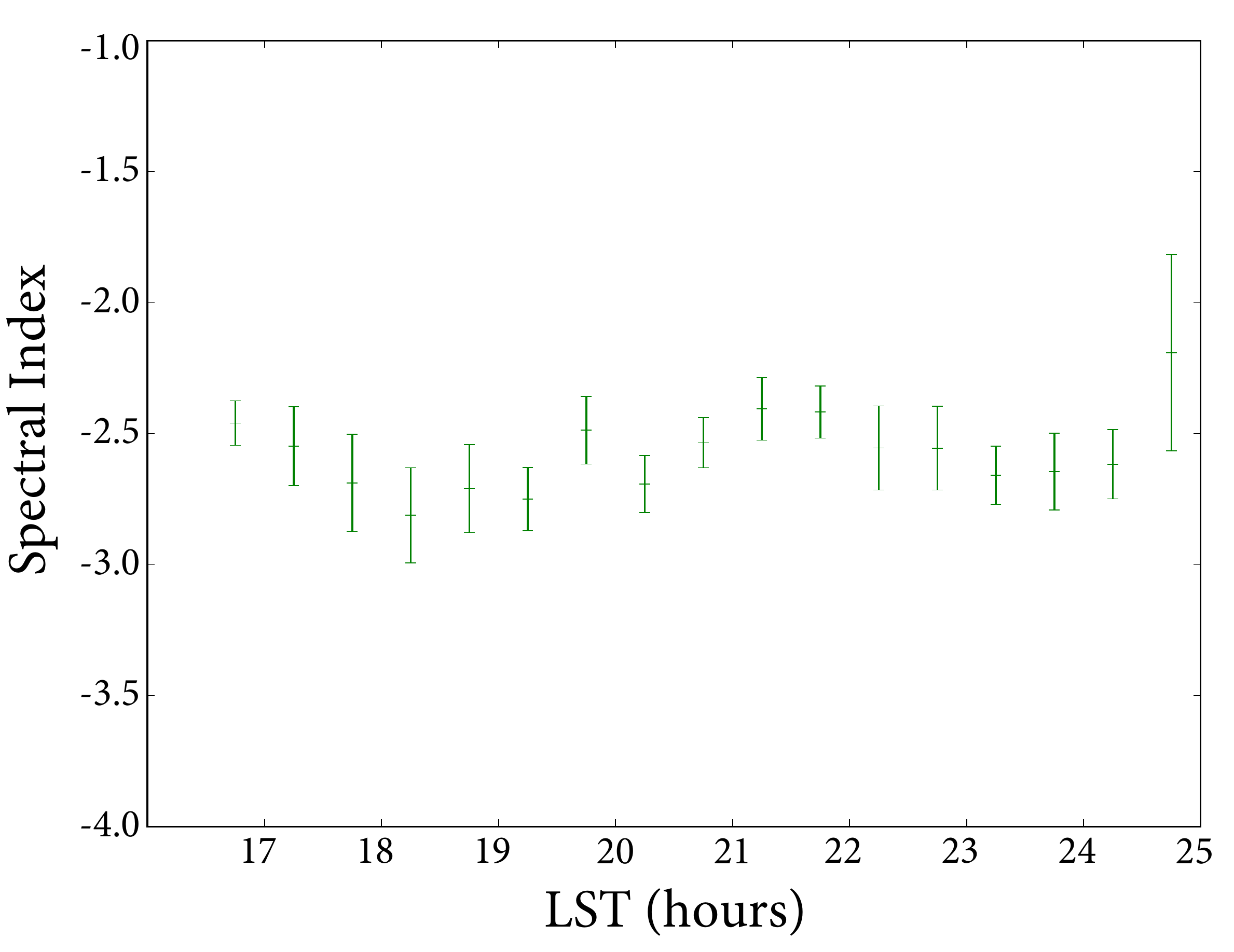}}
	\caption[Overall spectral index fit and beam-averaged spectral index over LST.]{
		Map-averaged overall spectral index fit (top) and beam-averaged spectral index over LST (bottom). The relative amplitudes and error bars in the top plot are obtained from the iterative procedure described in \refsec{subsecspectral}. For the bottom plot, each LST is observed on 5 different frequencies on 5 different nights (see \reffig{figmiteordata}), so we can obtain a spectral index at each LST by fitting the overall amplitude over frequency with a power law. The error bars are $1\sigma$, empirically estimated using the residuals of the power law fits.
		\label{figspectralindex}
	}
\end{figure}

In the iterative process to compute the 150\,MHz map, we also obtain relative amplitudes between all the data sets at different frequencies. At each frequency, we calculate visibilities using the 150\,MHz map we obtained as a model, and compute the relative amplitude by comparing them to our data. We then perform a linear fit in the log(amplitude)-log(frequency) space, and compute an overall spectral index of $-2.73\pm0.11$, as shown in \reffig{figspectralindex}. The error bar is calculated using the residuals in the fitting process. In addition to this overall spectral index averaged over the entire data set, we also perform the same procedure on subsets of the data, and fit for spectral indices for every half an hour of LST. The resulting time series of spectral indices varies smoothly between -2.4 and -2.8, as shown in \reffig{figspectralindex}.

In comparison to our spectral index result, the Experiment to Detect the Global EoR Signature (EDGES; \citealt{EDGES}) measured spectral index of $-2.5\pm0.1$ \citep{EDGESspectralindex}, centering at an out-of-plane region at declination $-26.5^\circ$ and right ascension 2\,h. Our overall spectral index agrees with the spectral index obtained by EDGES, with a $1.5\sigma$ difference. However, the difference is likely not due to statistical variation alone. The overall spectral index we present is averaged over the northern sky, so we are observing a different patch of sky compared to EDGES. Unlike EDGES, our result is influenced by both Cyg A and Cas A\footnote{The point sources are present in our spectral index results because our spectral index results are obtained in the visibility space, whereas the CLEAN algorithm that removed the point sources is performed in the image space.}. These strong radio sources have spectral indices of -2.7 and -2.8, respectively \citep{Parker68, Viny06, Viny14}, so they would shift our results towards a steeper spectral index. 

%------------------------------------------------------------
\subsubsection{Spectral Indices from 85\,MHz to 408\,MHz}
\begin{figure}
	\centerline{\includegraphics[width=.45\textwidth]{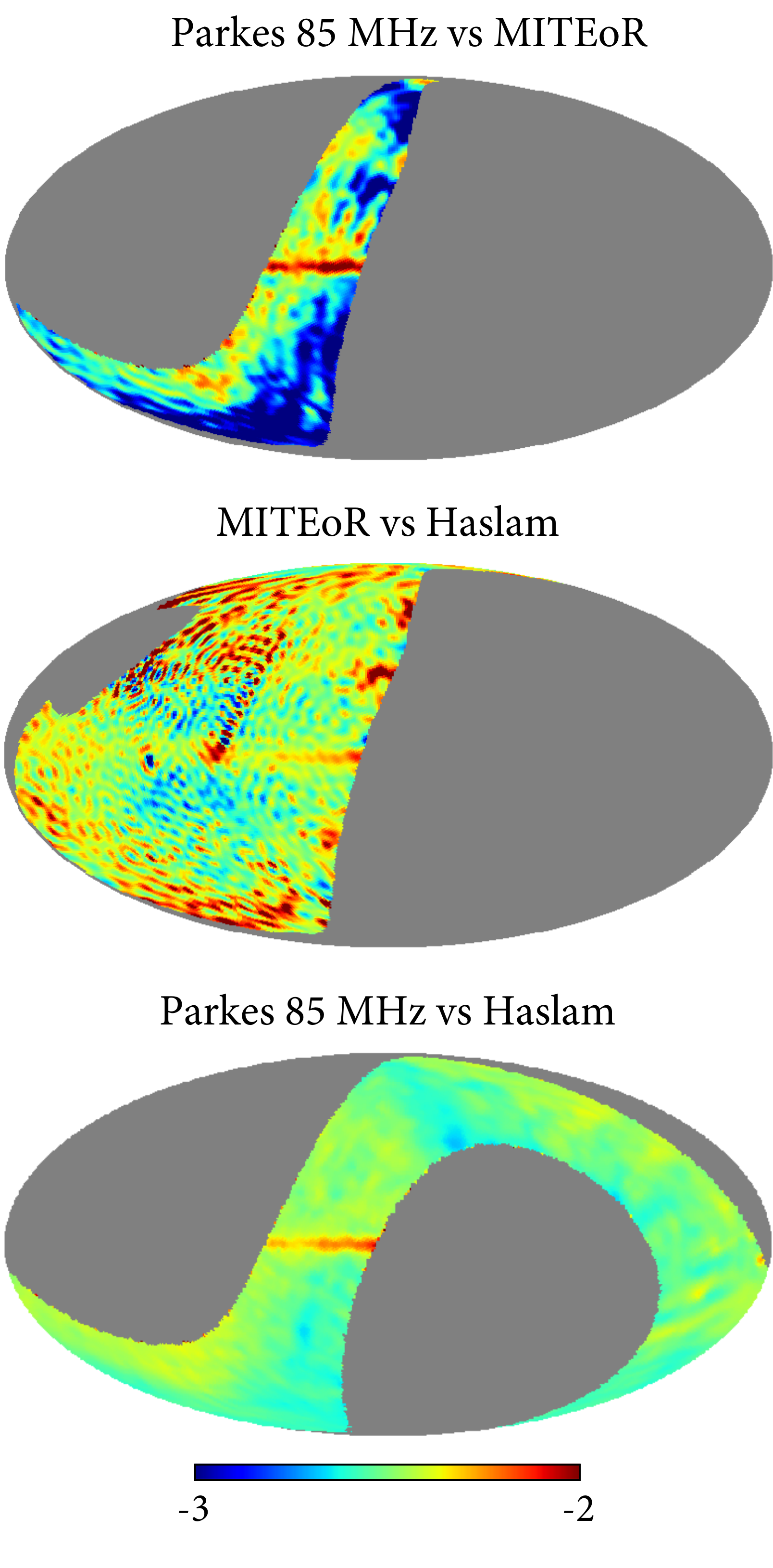}}
	\caption[Spectral index maps.]{
		Spectral index maps between the Parkes map at 85\,MHz, the MITEoR map at 150\,MHz, and the Haslam map at 408\,MHz. The MITEoR map is masked for regions with FWHM above $2.5^\circ$ or error bar above 20\,K. The median spectral indices in these maps are $-2.60$, $-2.43$, and $-2.50$, respectively. 
		\label{figspectralindexmaps}
	}
\end{figure}

In addition to spectral indices in the EoR frequency range computed using the MITEoR data alone, we also calculate maps of spectral indices by comparing our MITEoR map to the Parkes map at 85\,MHz \citep{Landecker1970Parkes} and the Haslam map at 408\,MHz \citep{Haslam1981, Haslam1982, Remazeilles2015Haslam}. We compute per-pixel spectral index maps for all three pairs of these three maps, as shown in \reffig{figspectralindexmaps}. 

The medians of the spectral indices shown in \reffig{figspectralindexmaps} are $-2.60\pm0.29\pm0.07$, $-2.43\pm0.18\pm0.04$, and $-2.50\pm0.07$, respectively. The first error bars come from the spread in spectral indices over the sky, and the second error bars come from 4\% absolute calibration uncertainty of MITEoR. This is a very weak indication that the spectral index softens over the range from 85\,MHz to 408\,MHz. For comparison, \citet{platania2003} presented an overall spectral index of $-2.695\pm0.120$ between the 408\,MHz \citep{Haslam1981, Haslam1982, Remazeilles2015Haslam}, 1.4\,GHz \citep{Reich1982Stockert, Reich1986Stockert, Reich2001Elisa}, and 2.3\,GHz \citep{Jonas1998Rhodes} maps, which also agrees well with an earlier study at these frequencies in \citet{Giardino2002}.

There are two spatial features worth noting in these maps. First, the Galactic plane has softer spectral indices than the out-of-plane regions. The median spectral indices within $5^\circ$ of the Galactic plane for the three map pairs are -2.27, -2.37, and -2.37, respectively. Softer spectral indices in the Galactic plane are also observed at higher frequencies in \citet{platania2003}, whose spectral index map comes from three maps above the EoR frequency range, as mentioned above.

In addition to softer indices in the Galactic plane, there are regions that clearly deviate from the median near the Galactic poles, such as the blue regions in the Parkes vs MITEoR map, and the red regions in the MITEoR vs Haslam map. Since such departure is not seen in the Parkes vs Haslam map, this suggests that the MITEoR map is about 50\,K lower in the Galactic pole regions than what the Parkes and Haslam maps jointly predict. There are two possible causes for this. The first is that the 50\,K deficiency in the MITEoR map is due to systematic errors. However, the temperatures in these regions are about 240\,K, so neither the 4\% absolute calibration uncertainty nor the $\sim15$\,K error bars can fully explain the 50\,K difference. Another possible cause is that the MITEoR map has a higher dynamic range than the Haslam and Parkes maps, so that it recovers more details at the low end of the temperature range compared to these maps. We leave a more careful investigation of this 50\,K discrepancy to a future study.

%%%%%%%%%%%%%%%%%%%%%%%%%%%%%
%%%%%%%%%%%%%%%%%%%%%%%%%%%%%
%%%%%%%%%%%%%%%%%%%%%%%%%%%%%
\section{Summary and Outlook}\label{secsummaryMM}
We have presented a new method for mapping diffuse sky emission using interferometric data. We have demonstrated its effectiveness through simulations for both MITEoR and the MWAcore, where we obtained maps with better than 50\,K noise and better than $2^\circ$ resolution for both instruments. We applied this method on the MITEoR data set collected in July 2013, which was absolutely calibrated using Cyg A and Cas A. We obtained a northern sky map averaged from 128\,MHz to 175\,MHz, with around $2^\circ$ angular resolution, 5\% uncertainty in its overall amplitude, and better than 100\,K noise. We also obtained an overall spectral index of $-2.69\pm0.11$, and beam-averaged spectral indices that vary over LST between -2.4 and -2.8. Both the MITEoR visibility data and the 150\,MHz sky map are available at \url{space.mit.edu/home/tegmark/omniscope.html}.

As this is our first application of this new method, there are many aspects of it that we are excited to investigate in future work. Throughout this work, we have focused on regularization matrices that are multiples of the identity matrix. However, since the sensitivity varies across the sky, especially in the case of MWAcore, a regularization matrix whose strength varies with sensitivity may achieve a better balance between noise suppression and PSF. It is also interesting to study what the optimal array layout is for imaging diffuse structure, along the lines of \citet{dillon_parsons2016}. Since the Earth rotates in the East-West direction, we expect the optimal array layout to be very different in the E-W direction than the N-S direction, perhaps similar to that of PAPER or CHIME \citep{CHIMEMM}. Moreover, it is interesting to investigate the effectiveness of this algorithm for instruments with much narrower primary beams, such as HERA. Lastly, it is valuable to perform further study in the effectiveness of this method for the purpose of calibration, such as calibrating the shorter baselines of MWA, which complements the existing calibration algorithms that focus on point source models for very long baselines.

{\bf Acknowledgments:}
MITEoR was supported by NSF grants AST-0908848, AST-1105835, and AST-1440343, the MIT Kavli Instrumentation Fund, the MIT undergraduate research opportunity (UROP) program, FPGA donations from XILINX, and by generous support from Jonathan Rothberg and an anonymous donor. AL acknowledges support from the University of California Office of the President Multicampus Research Programs and Initiatives through award MR-15-328388, and from NASA through Hubble Fellowship grant HST-HF2-51363.001-A awarded by the Space Telescope Science Institute, which is operated by the Association of Universities for Research in Astronomy, Inc., for NASA, under contract NAS5-26555. We wish to thank Jacqueline Hewitt and Aaron Parsons for helpful comments and suggestions, Evgenij Vinyajkin for his timely help with our error estimation for the Cyg A spectrum, Dan Werthimer and the CASPER group for developing and sharing their hardware and teaching us how to use it, Richard Armstrong, Matt Dexter, Alessio Magro, Mike Matejek, Qingxuan Pan, Robert Penna, Courtney Peterson, Meng Su, and Chris Williams for help with earlier stage of our hardware development and deployment.

%%%%%%%%%%%%%%%%%%%%%%%%%%%%%
%%%%%%%%%%%%%%%%%%%%%%%%%%%%%
\bibliographystyle{mnras}
\bibliography{MMGSM2}
%%%%%%%%%%%%%%%%%%%%%%%%%%%%%
%%%%%%%%%%%%%%%%%%%%%%%%%%%%%
%%%%%%%%%%%%%%%%%%%%%%%%%%%%%
%%%%%%%%%%%%%%%%%%%%%%%%%%%%%
%\newpage
\appendix

%%%%%%%%%%%%%%%%%%%%%%%%%%%%%
%%%%%%%%%%%%%%%%%%%%%%%%%%%%%
%%%%%%%%%%%%%%%%%%%%%%%%%%%%%
\section{Improvements to Redundant Calibration}\label{appdegen}
\citet{MITEoR} first demonstrated  the precision and speed of redundant calibration, which has since been applied to PAPER data analysis in the latest EoR power spectrum upper limits \citep{PAPERpspec2}. In this work we use the same core algorithms as described in \citet{MITEoR}, with improvements that make the algorithm much easier to use. In Appendix~\ref{apprough}, we describe the improvement to rough calibration, which is the first step in the redundant calibration process. A similar algorithm has been discussed and used in \citet{PAPERpspec}. In Appendix~\ref{appbreak}, we discuss how to reduce the number of undetermined quantities from two per time stamp to two per observing session. The completed redundant calibration software is publicly available at \url{https://github.com/jeffzhen/omnical}.

%%%%%%%%%%%%%%%%%%%%%%%%%%%%%
%%%%%%%%%%%%%%%%%%%%%%%%%%%%%
\subsection{Improved Rough Calibration}\label{apprough}
As described in \citet{MITEoR}, redundant calibration is a three step process: rough calibration, log calibration, and linear calibration. While log calibration and linear calibration do not rely on any sky information, they require rough calibration to get started. \citet{MITEoR} described a rough calibration algorithm that requires a sky model, which makes the whole pipeline rather cumbersome, especially when dealing with data sets from a new season or a new instrument. In this section we describe a new rough calibration algorithm that can be directly applied to data without any preprocessing, thus making the entire redundant calibration pipeline sky independent. The computational complexity is proportional to the number of antennas.

Here we describe our algorithm using an $4$ by $4$ array on a rectangular grid as shown in Table \ref{tabgrid}, but the algorithm is easily generalized to other redundant configurations. At a given time and frequency, we start with $16\times15/2=120$ visibilities, $v_{ij}$, whose phases are $\gamma_{ij}$, where $i,j$ are antenna numbers. The goal of rough calibration is to obtain antenna calibration phases, $\phi_i$, for all 16 antennas. The simple equation that describes phase calibration is 
\begin{equation}\label{eqphasecal}
\gamma_{ij} = -\phi_i + \phi_j + \theta_{i-j}
\end{equation}
where $\theta_{i-j}$ is the true phase shared by all redundant baselines that share the same baseline type with $\gamma_{ij}$, and we omit $2\pi$ wrapping for this section. 

We start by taking advantage of the three degree of freedom in phase degeneracies that are intrinsic to redundant calibration, namely an overall phase to all antenna phases, and two phases corresponding to rephasing the array. Due to these degeneracies, we are free to declare that $\phi_1 = 0$, $\theta_{1-2} = \gamma_{1-2}$, and $\theta_{1-5} = \gamma_{1-5}$ \footnote{We choose $\theta_{1-2}$ and $\theta_{1-5}$ because they are the two shortest non-parallel baseline types.}. This then gives us
\begin{equation}
\phi_1 = \phi_2 = \phi_5 = 0.
\end{equation}

With the first five phases $\phi_1, \phi_2, \phi_5, \theta_{1-2}, \theta_{1-5}$ all determined, we can now proceed to solve all 16 antenna phases. By applying \refeq{eqphasecal} on baseline $\gamma_{2,3}$, we can solve for $\phi_3$ since $\phi_2$ and $\theta_{2-3} = \theta_{1-2}$ are known. We can repeat the process to obtain $\phi_4$. Similarly, we can obtain $\phi_9$ using $\gamma_{5,9}$, and extend that to obtain $\phi_{13}$. Now we see that after determining the first five phases, solving for all the antennas is simply a matter of traversing the entire array and visiting every antenna with either one of the two baseline types we picked, in this case $\theta_{1-2}$ and $\theta_{1-5}$.

As shown in Table \ref{tabgrid}, the only baselines we used in this example are the arrows in the table, so the computational complexity is $\BigO(N)$ where $N$ is the number of antennas. If noise is a concern, one can increase the number of baseline types used to visit each antenna, so each phase calibration is sampled multiple times using multiple baseline types. For both MITEoR and PAPER 32 element data \citep{PAPERpspec}, we found that using just the two most redundant baseline types suffices.
\begin{table}
	\centering
	\begin{tabular}{  m{.6cm} m{.6cm}  m{.6cm} m{.6cm}  m{.6cm} m{.6cm} m{.6cm}}
		%\vskip2mm
		$\smile_1$ & $\longrightarrow$ & $\smile_2$ & $\longrightarrow$ & $\smile_3$ & $\longrightarrow$ & $\smile_4$ \\
		%\vskip2mm
		$\,\downarrow$ &  & $\,\downarrow$ &  & $\,\downarrow$ &  & $\,\downarrow$ \\
		%\vskip2mm
		$\smile_5$ &  & $\smile_6$ &  & $\smile_7$ &  & $\smile_8$ \\
		%\vskip2mm
		$\,\downarrow$ &  & $\,\downarrow$ &  & $\,\downarrow$ &  & $\,\downarrow$ \\
		%\vskip2mm
		$\smile_9$ &  & $\smile_{10}$ &  & $\smile_{11}$ &  & $\smile_{12}$ \\
		%\vskip2mm
		$\,\downarrow$ &  & $\,\downarrow$ &  & $\,\downarrow$ &  & $\,\downarrow$ \\
		%\vskip2mm
		$\smile_{13}$ &  & $\smile_{14}$ &  & $\smile_{15}$ &  & $\smile_{16}$ \\
		%\vskip2mm
		
	\end{tabular}
	\caption[A $4$ by $4$ antenna array on a regular grid.]{
		A $4$ by $4$ antenna array on a regular grid. Each $\smile_n$ represents an antenna, and each arrow represents a baseline used in the rough calibration algorithm.
		\label{tabgrid}
	}
\end{table}

It's important to note that while this rough calibration is much faster than log calibration and linear calibration, it is not advisable to re-run it on every data set before log calibration. As we pointed out above, rough calibration makes an arbitrary decision on the phase degeneracies, which later requires absolute calibration (with an accurate sky model) to determine. If each time step (typically a few seconds) uses a different rough calibration, we will have to run absolute calibration separately on every time step, which is very challenging as we may not always have good calibrators in the sky. It is much easier to run rough calibration only once at a single time stamp, and use it for a long period of data, such as one night or even a whole season. This way the entire time window will share one constant set of phase degeneracies, which are easy to determine in absolute calibration. Similarly, log calibration and linear calibration also make arbitrary choices of these same re-phasing degrees of freedom, so it's important to project out any re-phasing components from the calibration phase solutions, which we discuss in the next section.

%%%%%%%%%%%%%%%%%%%%%%%%%%%%%
%%%%%%%%%%%%%%%%%%%%%%%%%%%%%
\subsection{Removal of Temporal Fluctuation in Phase Calibration Degeneracies}\label{appbreak}
Both log calibration and linear calibration have the same set of phase degeneracies that cannot be determined by the self-consistency of the redundant measurements in a redundant array. Without careful treatment, these degeneracies can vary over time and become very difficult to determine. In this section we describe how we project out the temporal fluctuation in phase degeneracies in our redundant calibration solutions.

We start by reiterating the mathematical framework we first described in \citet{MITEoR}. When calibrating visibilities at any given time stamp and frequency bin, the quantity that log calibration minimizes is
\begin{align}\label{eqlogdegen}
\sum_{jk} |(\theta_{j-k}-\phi_j+\phi_k)- \gamma_{jk}|^2,
\end{align}
whereas linear calibration minimizes
\begin{align}\label{eqlindegen}
&\sum_{jk} |(y_{j-k}g_j^*g_k)-v_{jk}|^2\nonumber\\
=& \sum_{jk} \left||y_{j-k}g_j^*g_k|\exp\left[i(\theta_{j-k}-\phi_j+\phi_k)\right]-v_{jk}\right|^2.
\end{align}

For any given set of $\theta_{j-k}$ and $\phi_k$, one can add an arbitrary linear field $\boldsymbol{\mathnormal{\Phi}}\cdot\boldsymbol{r}_j+\psi$ to $\phi_j$ across the entire array, and simultaneously subtract $\boldsymbol{\mathnormal{\Phi}}\cdot\boldsymbol{d}_j$ from $\theta_{j-k}$, without changing the minimized quantities:
\begin{align}
\theta'_{j-k}-\phi'_j+\phi'_k \equiv& ( \theta_{j-k} - \boldsymbol{\mathnormal{\Phi}}\cdot\boldsymbol{d}_{j-k}) -(\phi_j+\boldsymbol{\mathnormal{\Phi}}\cdot\boldsymbol{r}_j+\psi) \nonumber \\ 
&+(\phi_k+\boldsymbol{\mathnormal{\Phi}}\cdot\boldsymbol{r}_k+\psi) \nonumber \\ 
=& \theta_{j-k}-\phi_j+\phi_k.
\end{align}
Here $\boldsymbol{d}_{j-k} \equiv \boldsymbol{r}_k-\boldsymbol{r}_j$ is the baseline vector for the unique baseline with best-fit visibility $y_{j-k}$. Thus, the quantities in expressions \ref{eqlogdegen} and \ref{eqlindegen} that the calibrations minimize are degenerate with changes to the linear phase field $\boldsymbol{\mathnormal{\Phi}}$ and the scalar $\psi$. This gives us 4 degenerate phase parameters: one overall phase $\psi$ (which has no impact on visibilities) and three related to the three degrees of freedom of the linear function $\boldsymbol{\mathnormal{\Phi}}$ (which reduces to two for a planar array). 

From the discussion above, we have two phase degeneracies per set of redundant visibilities at each time stamp that must be determined using other methods. Solving for these degeneracies at every time stamp can be very challenging, if not infeasible, given the large field of view of the instruments and limited availability of high quality calibrators. Therefore, we need to project out the temporal variability of the degeneracies so that we only have two degeneracies per observing session. To do this, we first pick a time stamp $t_0$ near the middle of the observing session, where the phase solutions for all antennas form a vector $\boldsymbol{\phi}_0$. Then, for any other time $t_1$ and solution $\boldsymbol{\phi}_1$, we project out the degeneracy difference between them by calculating
\begin{equation}\label{projectdegen}
\boldsymbol{\phi}'_1 = \mathbf{P}\boldsymbol{\phi}_0 + (\mathbf{I} - \mathbf{P})\boldsymbol{\phi}_1,
\end{equation}
where the projection matrix $\mathbf{P}$ into the degeneracy space is defined as 
\begin{equation}
\mathbf{P} \equiv \mathbf{R}(\mathbf{R}^t\mathbf{R})^{+}\mathbf{R}^t,
\end{equation}
and $\mathbf{R}$ is an $n_\text{ant}$ by 4 matrix whose $i$th row is $[\boldsymbol{r}_i\, 1]$, $\sum_i{\boldsymbol{r}_i}=\boldsymbol{0}$\footnote{Picking the origin for $\boldsymbol{r}_i$ to be center of the array simplifies the calculation in the next paragraph, with no impact on any visibility results.}, and ${}^+$ denotes pseudo-inverse. In this set-up, the phase degeneracy space is the column space of $\mathbf{R}$, and the degeneracy coefficients $(\boldsymbol{\mathnormal{\Phi}}, \psi)$ are the coordinates in this degeneracy space. Regardless of the choice of $t_1$, $\boldsymbol{\phi}'_1$ always has the same component in the degeneracy space, $\mathbf{P}\boldsymbol{\phi}_0$. We thus replace $\boldsymbol{\phi}_1$ by $\boldsymbol{\phi}'_1$ as our new phase calibration solution. 

While using $\mathbf{P}$ to project out degeneracies removes temporal variability in the degeneracies, it has the drawback of projecting out the `degeneracy component' in the variation of true phase solutions at the same time. Fortunately, since the projection is from a $n_\text{ant}$-dimensional space to a 2-dimensional space, the error caused by projection is much smaller than the phase drift in the calibration solutions.

To be more quantitative, let $\boldsymbol{\phi}_\text{var}$ be the difference vector in true phase calibration solutions between $t_0$ and $t_1$, $\boldsymbol{\phi}_\text{err}$ the part of $\boldsymbol{\phi}_\text{var}$ that is erroneously projected out, $\mathbf{R}_0$ an $n_\text{ant}$ by 4 matrix whose $i$th row is $[\boldsymbol{r}_i\, 0]$, and $\mathbf{P}_0 \equiv \mathbf{R}_0(\mathbf{R}^t\mathbf{R})^{+}\mathbf{R}^t$\footnote{When $(\mathbf{R}^t \mathbf{R})^+ \mathbf{R}^t$ is applied to the calibration solutions, it gives the degeneracy coefficients $(\boldsymbol{\mathnormal{\Phi}}, \psi)$. We can use these coefficients to convert back to the actual degenerate components of the solution, but we use $\mathbf{R}_0$ rather than $\mathbf{R}$ because we do not want to penalize ourselves for the overall degeneracy $\psi$ that does not impact visibilities.}. We then have
\begin{equation}
\boldsymbol{\phi}_\text{err} = \mathbf{P}_0\boldsymbol{\phi}_\text{var},
\end{equation}
and
\begin{equation}
\langle\boldsymbol{\phi}_\text{err}\boldsymbol{\phi}^t_\text{err}\rangle = \mathbf{P}_0\langle\boldsymbol{\phi}_\text{var}\boldsymbol{\phi}^t_\text{var}\rangle\mathbf{P}^t_0.
\end{equation}

Assuming the drift for each antenna is uncorrelated and has a typical magnitude of $\epsilon$, the typical magnitude of phase projection error can be described by the median of the diagonal entries of $\epsilon\sqrt{\mathbf{P}_0\mathbf{P}^t_0}$. For an 8 by 8 regular grid array such as MITEoR, median of the diagonal of $\sqrt{\mathbf{P}_0\mathbf{P}^t_0}$ is 17.6\%, and for a 16 by 16 array this becomes 8.8\%, for a 32 by 32 array 4.3\%. This result is unsurprising, since we would expect the error fraction to be roughly $\sqrt{2/n_\text{ant}}$ using the dimensionality argument alone.
Therefore, the error caused by the projection operation diminishes as the number of antennas grows.

Lastly, an important subtlety worth noting is phase wrapping on $(-\pi, \pi]$. All the equations in this section are dealing with phase, so they are subject to phase wrapping. Phase wrapping can potentially render \refeq{projectdegen} invalid. Therefore, rather than calculating \refeq{projectdegen} directly, it is much less likely to have phase wrapping by calculating
\begin{equation}\label{projectdegen2}
\boldsymbol{\phi}'_1 = \boldsymbol{\phi}_0 + (\mathbf{I} - \mathbf{P})(\boldsymbol{\phi}_1 - \boldsymbol{\phi}_0).
\end{equation}
Unlike $\boldsymbol{\phi}_0$ and $\boldsymbol{\phi}_1$ which can have any values from $-\pi$ to $\pi$, $\boldsymbol{\phi}_1 - \boldsymbol{\phi}_0$ should be small angles around 0. For large arrays, even $\boldsymbol{\phi}_1 - \boldsymbol{\phi}_0$ will have large angles that wrap around $\pi$ for antennas near the outer part of the array. In this case, one can first take only the inner part of the array to find a rough solution to the degeneracy coefficients, and remove them from them from $\boldsymbol{\phi}_1 - \boldsymbol{\phi}_0$ to reduce the angles therein and eliminate phase wrapping, and then carry out \refeq{projectdegen2}.

%%%%%%%%%%%%%%%%%%%%%%%%%%%%%
%%%%%%%%%%%%%%%%%%%%%%%%%%%%%
%%%%%%%%%%%%%%%%%%%%%%%%%%%%%
\section{Dynamic Pixelization}\label{appdynamic}
 Multiplications and inversions of $10^5$ by $10^5$ matrices can currently be done in a matter of days, but since the computational times grows as $\mathcal{O}(\Delta\Omega^{-3})$, improving the resolution can be very computationally demanding. To reduce the matrix size. we can resort to a more intelligent pixelization scheme, which we call dynamic pixelization.
 
 Having uniform pixelization can be wasteful. For example, near the edge of our observable region, fine pixelization brings little advantage since our sensitivity is very low. In contrast, regions with strong flux demand much finer pixelization, such as the Galactic center and strong point sources such as Cas A and Cyg A. We use a recursive algorithm to generate a pixelization scheme that accommodates all of the above considerations. We start with a map predicted by the GSM\footnote{Note that using the GSM for pixelization does not automatically mix GSM information into our solution.}, and a uniform but coarse HEALPIX nested pixelization with $n_\text{side}\sim n_\lambda$ for the longest baseline\footnote{This relation is derived from $\frac{4\pi}{n_\text{pix}}\sim\left(\frac{1}{n_\lambda}\right)^2$.}. We then calculate a beam weighted sky map $s_\text{bweight}$ given by
 \begin{equation}
 s_\text{bweight}(\hat{\bm{k}}) = \int_t{s(\hat{\bm{k}})B(\hat{\bm{k}}, t)dt}.
 \end{equation}
 Lastly, we recursively divide each pixel into 4 sub-pixels according to HEALPIX nest scheme, until the standard deviation among the $s_\text{bweight}$ values at these 4 sub-pixels are below a threshold value $\sigma$. Intuitively, since each pixel corresponds to a column in the $\A$-matrix, the algorithm only splits each column into 4 when the resulting 4 columns will be sufficiently different in fitting the data.
 
\begin{figure}
 	\centerline{\includegraphics[width=.4\textwidth]{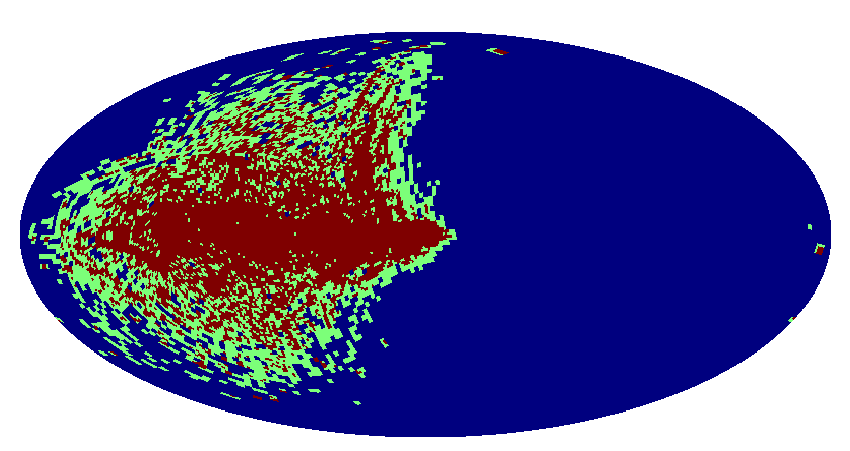}}
 	\centerline{\includegraphics[width=.45\textwidth]{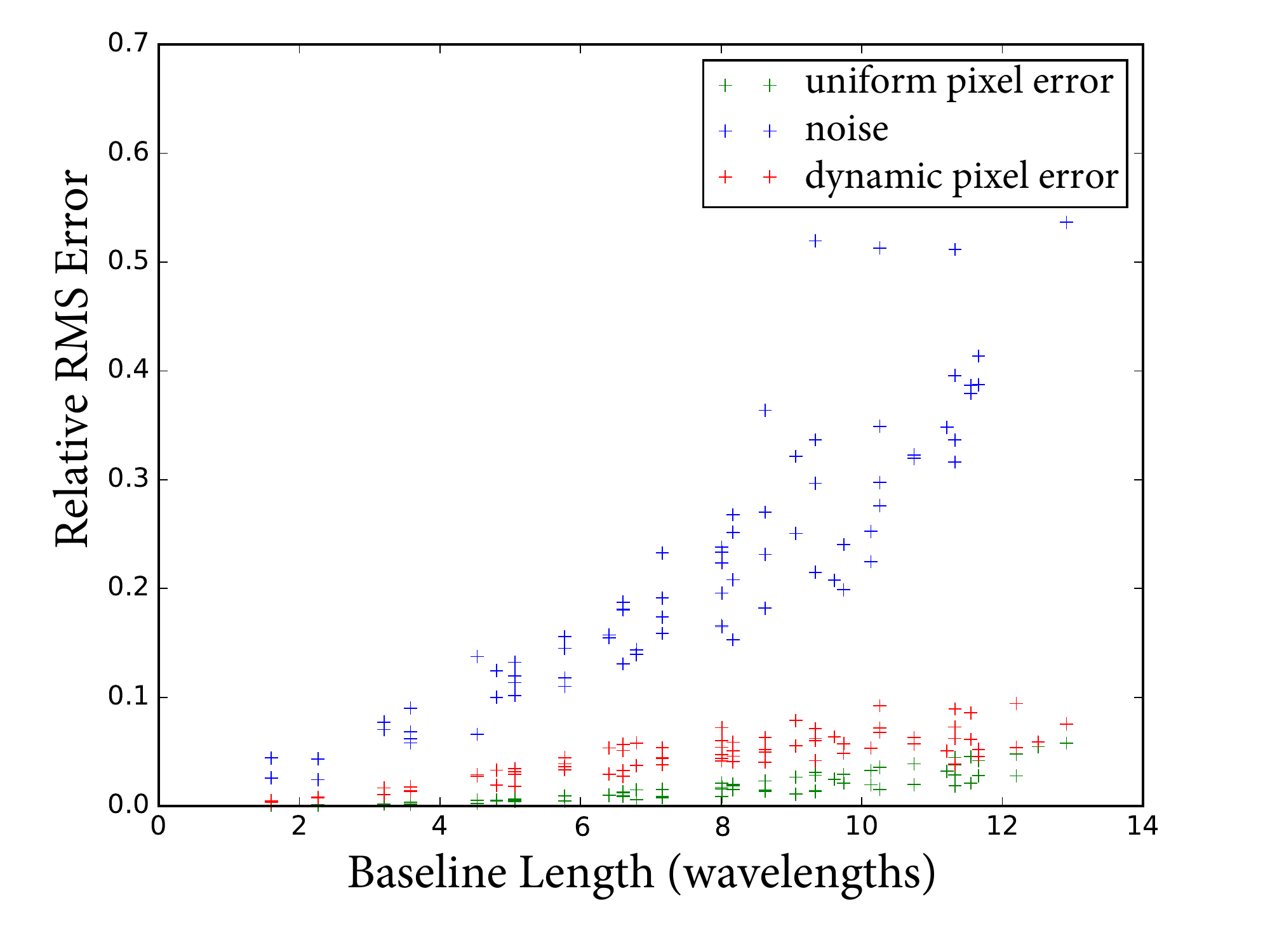}}
 	\caption[Dynamic pixelization scheme.]{
 	The upper plot shows a dynamic pixelization scheme, where the blue, green, and red regions represent $n_\text{side}$ of 32, 64, and 128, respectively. This scheme contains 48900 total pixels, 24.9\% of the pixel count in a uniform $n_\text{side} = 128$ pixelization. The lower plot shows the errors in the simulated visibilities compared to those simulated with a $n_\text{side} = 512$ pixelization, and we see that the dynamic pixelization error is a factor of two higher than uniform pixelization, and it is negligible compared to the noise level assuming one night's observation. 
 	}
 	\label{figpixelerror}
\end{figure}

 We decide the threshold $\sigma$ by numerically searching for the optimal choice. As larger $\sigma$ translates into fewer pixels but less pixel precision, we choose the largest $\sigma$ such that the simulated pixelization error is less than $1\%$ in power compared to the noise in the visibility data. \reffig{figpixelerror} shows an example pixelization scheme for MITEoR data at 160\,MHz, where the final pixel count is less than a quarter of uniform pixelization, leading to a factor of 64 speed-up in computation time, while the pixelization error remains negligible compared to noise in the data.

%%%%%%%%%%%%%%%%%%%%%%%%%%%%%
%%%%%%%%%%%%%%%%%%%%%%%%%%%%%
%%%%%%%%%%%%%%%%%%%%%%%%%%%%%
%\section{Appendix: $(\X^{-1} + \Y^{-1})^{-1} = \X(\X + \Y)^{-1}\Y$}\label{appXY}
%\begin{proof}
%\begin{align}
%&(\X^{-1} + \Y^{-1})^{-1} = \X(\X + \Y)^{-1}\Y\\
%\Leftrightarrow&(\X^{-1} + \Y^{-1})\X(\X + \Y)^{-1}\Y = \I\\
%\Leftrightarrow&(\I + \Y^{-1}\X)(\X + \Y)^{-1}\Y = \I\\
%\Leftrightarrow&\Y^{-1}(\Y + \X)(\X + \Y)^{-1}\Y = \I\\
%\Leftrightarrow&\I  = \I\qedhere
%\end{align}
%
%\end{proof}

\end{document}

%% file: macros.tex
\newcommand{\refeq}[1]{Eq.~\eqref{#1}}
\newcommand{\refsec}[1]{Section \ref{#1}}
\newcommand{\refapp}[1]{Appendix \ref{#1}}
\newcommand{\reffig}[1]{Fig.~\ref{#1}}

\newcommand{\R}{\mathbf{R}}

\newcommand{\N}{\mathbf{N}}

\newcommand{\I}{\mathbf{I}}

\newcommand{\BigO}{\mathcal{O}}

\newcommand{\X}{\mathbf{X}}
\newcommand{\Y}{\mathbf{Y}}

\newcommand{\beq}{\begin{equation}}
\newcommand{\eeq}{\end{equation}}

\newcommand{\A}{\mathbf{A}}

\newcommand{\PSF}{\mathbf{P}}

\newcommand{\W}{\mathbf{W}}

\def\spose#1{\hbox to 0pt{#1\hss}}
\def\simlt{\mathrel{\spose{\lower 3pt\hbox{$\mathchar"218$}}
   \raise 2.0pt\hbox{$\mathchar"13C$}}}
\def\simgt{\mathrel{\spose{\lower 3pt\hbox{$\mathchar"218$}}
     \raise 2.0pt\hbox{$\mathchar"13E$}}}
 \def\simpropto{\mathrel{\spose{\lower 3pt\hbox{$\mathchar"218$}}
     \raise 2.0pt\hbox{$\propto$}}}

\newcommand{\be}{\begin{equation}}
\newcommand{\ee}{\end{equation}}

\def\Let@{\def\\{\notag\math@cr}}

\newcolumntype{L}[1]{>{\raggedright\let\newline\\\arraybackslash\hspace{0pt}}m{#1}}
\newcolumntype{C}[1]{>{\centering\let\newline\\\arraybackslash\hspace{0pt}}m{#1}}
\newcolumntype{R}[1]{>{\raggedleft\let\newline\\\arraybackslash\hspace{0pt}}m{#1}}

\newcount\colveccount
\newcommand*\colvec[1]{
        \global\colveccount#1
        \begin{pmatrix}
        \colvecnext
}
\def\colvecnext#1{
        #1
        \global\advance\colveccount-1
        \ifnum\colveccount>0
                \\
                \expandafter\colvecnext
        \else
                \end{pmatrix}
        \fi
}